\documentclass[prd,amsmath,amsfonts,floatfix,nofootinbib,superscriptaddress,preprintnumbers]{revtex4}
\usepackage{graphicx,bm}
\usepackage{amsmath,amssymb,amsbsy,mathbbol}
\usepackage{comment}

\newcommand{\epp}{\epsilon^\prime}
\newcommand{\xpt}[0]{{$\chi$PT}}
\newcommand{\qxpt}[0]{{Q$\chi$PT}}
\newcommand{\pqxpt}[0]{{PQ$\chi$PT}}
\newcommand{\hbxpt}[0]{{HB$\chi$PT}}
\newcommand{\str}[0]{\rm str}
\DeclareMathOperator{\tr}{tr}

\newcommand{\veps}{{\vec\epsilon}}
\newcommand{\vepsprime}{{\vec\epsilon\, '}}
\newcommand{\vsigma}{{\vec\sigma}}

\begin{document}

\def\a{{\alpha}}
\def\b{{\beta}}
\def\d{{\delta}}
\def\D{{\Delta}}
\def\e{{\epsilon}}
\def\g{{\gamma}}
\def\G{{\Gamma}}
\def\k{{\kappa}}
\def\l{{\lambda}}
\def\L{{\Lambda}}
\def\m{{\mu}}
\def\n{{\nu}}
\def\w{{\omega}}
\def\O{{\Omega}}
\def\S{{\Sigma}}
\def\s{{\sigma}}
\def\t{{\tau}}
\def\th{{\theta}}
\def\x{{\xi}}

\def\ol#1{{\overline{#1}}}

\def\Dslash{D\hskip-0.65em /}
\def\dslash{{\partial\hskip-0.5em /}}
\def\vslash{{\rlap \slash v}}
\def\qbar{{\overline q}}

\def\CPT{{$\chi$PT}}
\def\QCPT{{Q$\chi$PT}}
\def\PQCPT{{PQ$\chi$PT}}
\def\tr{\text{tr}}
\def\str{\text{str}}
\def\diag{\text{diag}}
\def\order{{\mathcal O}}
\def\vit{{\it v}}
\def\vD{\vit\cdot D}
\def\am{\alpha_M}
\def\bm{\beta_M}
\def\gm{\gamma_M}
\def\smb{\sigma_M}
\def\smt{\overline{\sigma}_M}
\def\tb{{\tilde b}}

\def\c#1{{\mathcal #1}}

\def\Bbar{\overline{B}}
\def\Tbar{\overline{T}}
\def\cBbar{\overline{\cal B}}
\def\cTbar{\overline{\cal T}}
\def\pq{(PQ)}

\def\eqref#1{{(\ref{#1})}}

\def\a{{\alpha}}
\def\b{{\beta}}
\def\d{{\delta}}
\def\D{{\Delta}}
\def\e{{\varepsilon}}
\def\g{{\gamma}}
\def\G{{\Gamma}}
\def\k{{\kappa}}
\def\l{{\lambda}}
\def\L{{\Lambda}}
\def\m{{\mu}}
\def\n{{\nu}}
\def\o{{\omega}}
\def\O{{\Omega}}
\def\S{{\Sigma}}
\def\s{{\sigma}}
\def\th{{\theta}}
\def\x{{\xi}}
\def\p{{\partial}}
\def\ol#1{{\overline{#1}}}
\def\cO{{\mathcal{O}}}
\def\c{{\chi}}
\def\cD{{\mathcal{D}}}
\def\cA{{\mathcal{A}}}
\def\cQ{{\mathcal{Q}}}

\preprint{NT@UW 05-12}

\title{Electromagnetic and spin polarisabilities in lattice QCD}

\author{W. Detmold}
\email[]{wdetmold@phys.washington.edu}
\affiliation{Department of Physics, University of Washington,
        Box 351560,
        Seattle, WA 98195-1560, USA}

\author{B. C. Tiburzi}
\email[]{bctiburz@phy.duke.edu}
\affiliation{Department of Physics,
Duke University,
P.O.~Box 90305,
Durham, NC 27708-0305, USA}

\author{A. Walker-Loud}
\email[]{walkloud@u.washington.edu}
\affiliation{Department of Physics, University of Washington,
        Box 351560,
        Seattle, WA 98195-1560, USA}

\date{\today}

\begin{abstract}
  We discuss the extraction of the electromagnetic and spin
  polarisabilities of nucleons from lattice QCD. We show that the
  external field method can be used to measure all the electromagnetic
  and spin polarisabilities including those of charged particles.  We
  then turn to the extrapolations required to connect such
  calculations to experiment in the context of finite volume chiral
  perturbation theory. We derive results relevant for lattice
  simulations of QCD, partially-quenched QCD and quenched QCD. Our
  results for the polarisabilities show a strong dependence on the
  lattice volume and quark masses, typically differing from the
  infinite volume limit by $\sim10$\% for current lattice volumes and
  quark masses.
\end{abstract}
\maketitle

%
%
%
%
%
%

\section{Introduction}

Compton scattering at low energies is an invaluable tool with which to
study the electromagnetic structure of hadrons. At very low photon
energies, the Compton amplitude is dominated by point-like photon
scattering from the total charge and magnetic moment of the target
hadrons. As the frequency increases, contributions beyond point-like
scattering enter and one begins to resolve the hadronic response to an
applied electromagnetic field.  For unpolarised scattering on spin
one-half objects, the first structure dependent contributions in this
energy expansion of the amplitude are the electric polarisability,
$\alpha$, and the magnetic polarisability, $\beta$. These quantities
reflect the ability of the hadron's components to align or anti-align
themselves in response to an applied electric or magnetic field. For
the proton and neutron, the positivity of the accepted experimental
values of these polarisabilities ($\alpha_p=12.0\pm0.6$,
$\beta_p=1.9\mp0.6$, $\alpha_n=12.5\pm1.7$ and $\beta_n=2.7\mp1.8$ in
units of $10^{-4}$ fm$^3$~\cite{Schumacher:2005an}) indicates that
both nucleons are diamagnetic objects.  Recent experimental
advances~\cite{Hyde-Wright:2004gh,Schumacher:2005an} have also allowed
the extraction of certain combinations of target
polarisation-dependent observables in Compton scattering. These
involve the so-called spin polarisabilities~\cite{Ragusa:1993rm},
conventionally labeled $\gamma_1$--$\gamma_4$, and they have
consequently been investigated in numerous theoretical and further
experimental studies. Although the classical interpretation of
spin-dependent Compton scattering is less clear, the spin
polarisabilities encode additional fundamental properties of the
nucleon.  Compton scattering observables, however, are not limited to
these six parameters. Higher order quasi-static properties of the
nucleon appear from further terms in the energy expansion of the
amplitude. These higher-order polarisabilities \cite{Holstein:1999uu},
as well as generalised polarisabilities \cite{Guichon:1995pu} (which
arise in the singly (doubly) virtual Compton scattering process,
$\gamma^* X \to \gamma^{(*)} X$) allow for an even finer resolution of
the electromagnetic structure of hadrons at low energies.

While experimentally one is hoping to open further windows through
which to view hadronic electromagnetic structure, theoretically one
ultimately hopes to understand how hadronic polarisabilities arise
from the basic electromagnetic interaction of the photon with quarks
that are bound to form the hadrons. The electric and magnetic
polarisabilities should na{\"\i}vely scale with the volume of the
hadron. However, this expectation overestimates the observed
polarisabilities by four orders of magnitude, indicating that the
nucleon's constituents are strongly coupled. Lattice techniques
provide a method to investigate the non-perturbative structure of
hadrons directly from QCD.  In particular, the various hadron
polarisabilities can be computed.  Comparison of these results with
experimental determinations would provide stringent tests of the
lattice method's ability to reproduce the structure of physical
hadronic states; for the individual spin polarisabilities that have
not been measured, the lattice approach may be the only way to
determine them. On the lattice, direct calculations of the required
hadronic current-current correlators are difficult and so far have not
been attempted. However significant progress has been made
\cite{Fiebig:1988en,Christensen:2004ca,Lee:2005dq} in extracting the
electric and magnetic polarisabilities by performing quenched lattice
calculations in constant background electric and magnetic fields
respectively and studying the quadratic shift in the hadron mass that
is induced (essentially an application of the Feynman-Hellman
theorem). These studies have investigated the electric
polarisabilities of various neutral hadrons (in particular, the
uncharged vector mesons and uncharged octet and decuplet baryons), and
the magnetic polarisabilities of the baryon octet and decuplet, as
well as those of the non-singlet pseudo-scalar and vector mesons.  As
we shall discuss below, generalisations of these methods using
non-constant fields allow the extraction of the spin polarisabilities
from spin-dependent correlation functions and also allow the electric
polarisabilities to be determined for charged hadrons. More generally,
higher-order polarisabilities and generalised polarisabilities are
accessible using this technique.

As with all current lattice results, these calculations have a number
of limitations and so are not physical predictions that can be
directly compared to experiment. For the foreseeable future, lattice
QCD calculations will necessarily use quark masses that are larger
than those in nature because of limitations in the available
algorithms and computational power. Additionally, the volumes and
lattice-spacings used in these calculations will always be finite and
non-vanishing, respectively.  For sufficiently small masses and large
volumes, the effects of these approximations can be investigated
systematically using the effective field theory of the low energy
dynamics of QCD, chiral perturbation theory ($\chi$PT)
\cite{Weinberg:1966kf,Gasser:1983yg,Gasser:1985gg}.\footnote{The
  effects of the lattice discretisation are short distance in nature,
  and while some of them can be analysed in an extension of the
  effective field theory described here
  \cite{Sharpe:1998xm,Rupak:2002sm,Bar:2003mh,Beane:2003xv}, others
  are not accounted for. Here we will assume that a continuum
  extrapolation has been performed.}  In this paper we shall perform
an analysis of the nucleon electromagnetic and spin polarisabilities
at next-to-leading order (NLO) in the chiral expansion.  We do so to
discuss the infrared effects of the quark masses and finite volume in
two-flavour QCD and its quenched and partially-quenched analogues
(QQCD and PQQCD).  The polarisabilities of the hadrons are
particularly interesting in this regard since they are very sensitive
to infrared physics and their quark mass and finite volume dependence
is considerably stronger than that expected for hadron masses and
magnetic moments.  This should be physically evident given that the
polarisabilities scale with the volume.  In essence, chiral
perturbation theory provides a model independent analysis of the
modification of the nucleon's pion cloud in a finite volume. When the
charged pion cloud is influenced by to the periodic boundary
conditions imposed on the lattice, the nucleon's response to external
electromagnetic fields is altered compared to that at infinite volume,
and in most cases the effects are dramatic.  A particularly striking
oddity that we find in this analysis is a modification of the Thompson
cross section at finite volume.  This can be explained through the
physics of chiral loop corrections to point-like hadron structure.

If future lattice QCD simulations are to provide physical predictions
for the electromagnetic and spin polarisabilities, careful attention
must be paid to both the chiral and infinite volume extrapolations. To
illustrate this point, we present our results at representative values
of the quark mass, finding significant effects.  We also use our
quenched chiral perturbation theory results to assess the volume
dependence of the quenched data at the lightest pion masses used in
Refs.~\cite{Christensen:2004ca,Lee:2005dq}.  While the quenched theory
contains unphysical low energy constants (LECs) and the convergence of
the chiral expansion is questionable at these pion masses, we can
still provide an estimate of the volume dependence of quenched data
for the nucleon polarisabilities using our results.  Such an estimate
is achievable because the corresponding polarisabilities in the
unquenched theory do not depend on phenomenologically undetermined
LECs at the order of the chiral expansion to which we work.  At the
lightest quark masses used in the existing quenched lattice
simulations, $m_\pi\sim0.5$~GeV, we find strong sensitivity to the
lattice volume (as large as 10\%).  The effects will only increase as
the pion mass is brought closer to that in nature.  Clearly careful
chiral and volume extrapolations of polarisabilities are mandated to
connect lattice calculations to real world QCD.

To begin our investigation of nucleon polarisabilities in lattice QCD,
we discuss in Sec.~\ref{sec:gN_to_gN} the kinematics of Compton
scattering and define the electromagnetic and spin polarisabilities
that are the primary focus of this work.  In Sec.~\ref{sec:Compton},
we perform a general analysis of the external field method pertaining
to all electromagnetic and spin polarisabilities.  We discuss how
suitable background fields can be used in lattice QCD simulations to
determine the spin polarisabilities and, more generally, generalised
polarisabilities (though we limit our discussion of these in the
present paper). Following this we introduce the low-energy effective
theories of QCD ($\chi$PT), quenched QCD (Q$\chi$PT) and
partially-quenched QCD (PQ$\chi$PT). These effective theories provide
the model independent input necessary for calculating the quark mass
and lattice volume dependence of polarisabilities. We focus primarily
on PQ$\chi$PT in Sec.~\ref{sec:hbxpt}, discussing the relation to
$\chi$PT where relevant, and relegating the peculiarities of Q$\chi$PT
to Appendix~\ref{A4}.  Our results for the dependence of the nucleon
polarisabilities on quark masses and the lattice volume are presented
in Sec.~\ref{sec:pols}.  We provide detailed plots relevant for full
QCD simulations of polarisabilities showing the dependence on quark
masses and lattice volumes. We also estimate the quenched QCD volume
dependences of the polarisabilities at a pion mass typical of existing
quenched lattice data.  Very small volumes are discussed in
Appendix~\ref{epp_regime} and a glossary of finite volume functions
required to evaluate the polarisabilities in a periodic box appears in
Appendix~\ref{FV_app}.  Lastly, Sec.~\ref{sec:discussion} consists of
a concluding discussion of our results.

%
%
%
%
%
%
%
%
%
%
%
%
\section{Compton scattering on spin-half hadrons}
\label{sec:gN_to_gN}

The real Compton scattering amplitude describing the elastic
scattering of a photon on a spin-half target such as the proton or
neutron can be parameterised as
\begin{eqnarray}
&&T_{\gamma N}=  A_1(\omega,\theta)\, \vepsprime\cdot\veps
            +A_2(\omega,\theta)\, \vepsprime\cdot \hat{k} \, \veps\cdot \hat{k'}
             +i\,A_3(\omega,\theta)\, \vsigma\cdot (\vepsprime\times\veps)
             +i\,A_4(\omega,\theta)\, \vsigma\cdot
             (\hat{k'}\times\hat{k})\, \vepsprime\cdot\veps 
     \nonumber \\
 & & 
\hspace*{1.2cm}
     +i\,A_5(\omega,\theta)\, \vsigma\cdot \left[(\vepsprime\times\hat{k})\,
     \veps\cdot\hat{k'} 
            -(\veps\times\hat{k'})\, \vepsprime\cdot\hat{k}\right]
     +i\,A_6(\omega,\theta)\, \vsigma\cdot \left[(\vepsprime\times\hat{k'})\,
     \veps\cdot\hat{k'} 
                        -(\veps\times\hat{k})\, \vepsprime\cdot\hat{k}\right],
\label{eq:Ti}
\end{eqnarray}
where we have chosen to work in the Breit frame of the system and the
incoming and outgoing photons have momenta
$k=(\omega,\vec{k}=\omega\,\hat{k})$ and
$k^\prime=(\omega,\vec{k}^\prime=\omega\,\hat{k}^\prime)$, and
polarisation vectors $\epsilon$ and $\epsilon^\prime$, respectively.
The $A_i(\omega,\theta)$, $i=1\ldots 6$, are scalar functions of the
photon energy and scattering angle,
$\cos\theta=\hat{k}\cdot\hat{k}^\prime$.  It is convenient to work in
Coulomb gauge throughout where $\epsilon_0=\epsilon_0^\prime=0$ (the
physical amplitudes are gauge invariant).

The functions, $A_i$, determining the Compton amplitude can be
separated into a number of pieces. The Born terms describe the
interaction of the photon with a point-like target with mass, $M_N$,
charge, $e\, Z$ (where $e>0$), and magnetic moment, $\mu$.  These
terms reproduce the Thomson-limit and quadratic frequency pieces
\cite{RaiChoudhury68} of unpolarised scattering and the
Low--Gell-Mann--Goldberger low energy theorems
\cite{Gell-Mann:1954kc,Low:1954kd} for spin-dependent scattering. The
remaining parts of the amplitude describe the structural response of the target.
Expanding the amplitude for small photon energies relative to the
target mass and keeping terms to ${\cal O}(\omega^3)$ one can write
\begin{eqnarray}
A_1(\omega,\theta) &=& -Z^2 \frac{e^2}{M_N}+\frac{e^2}{4M_N^3}
\left(\mu^2(1+\cos\theta)-Z^2\right)(1-\cos\theta)\,\omega^2
+ 4\pi(\alpha + \beta \, \cos\theta)\omega^2+
{\cal O}(\omega^4)\,,
\nonumber\\ 
A_2(\omega,\theta) &=& \frac{e^2}{4M_N^3} (\mu^2-Z^2) \omega^2
\cos\theta -4\pi\beta \omega^2 \,+ 
{\cal O}(\omega^4)\,,
\nonumber\\ 
A_3(\omega,\theta) &=&   \frac{e^2 \omega}{
2M_N^2}\left(Z(2\mu-Z)-\mu^2 \cos\theta\right) 
+ 4\pi\omega^3(\gamma_1 - (\gamma_2 + 2 \gamma_4) \, \cos\theta)+
{\cal O}(\omega^5)\,,
\nonumber\\ 
A_4(\omega,\theta) &=&  -\frac{e^2 \omega }{ 2
M_N^2 }\mu^2 + 4\pi\omega^3 \gamma_2 +
{\cal O}(\omega^5)\,,
\nonumber\\ 
A_5(\omega,\theta) &=&  \frac{e^2 \omega }{2 M_N^2 }\mu^2 +
4\pi\omega^3\gamma_4 + {\cal O}(\omega^5) \,,
\nonumber\\ 
A_6(\omega,\theta) &=&  -\frac{e^2 \omega}{2 M_N^2 } Z\mu +
4\pi\omega^3\gamma_3 + {\cal O}(\omega^5)\,,
\label{eq:amplitudes}
\end{eqnarray}
describing the target structure in terms of the electric, magnetic and
four spin polarisabilities, $\alpha$, $\beta$, and
$\gamma_{1\mbox{--}4}$, respectively. In the conventions above, the
spin polarisabilities receive contributions from the anomalous decay
$\pi^0\to\gamma\gamma$ (shown in Fig.~\ref{fig:anom} below). This
contribution varies rapidly with energy and is omitted from the
polarisabilities in some conventions. Higher order terms in the energy
expansion can be parameterised in terms of higher-order
polarisabilities \cite{Holstein:1999uu}. The more general process of
virtual (and doubly-virtual) Compton scattering at low energies can
similarly be described in terms of generalised
polarisabilities~\cite{Guichon:1995pu}. We will focus in the six
polarisabilities defined above.

The goal of this paper is to determine the quark mass and volume
dependence of the polarisabilities defined above to allow accurate
extraction of their physical values from lattice calculations.  Before
we do this we shall discuss how these lattice calculations may be
implemented.

%
%
%
%
%
%
%
%
%
%
%
%
\section{Compton scattering and polarisabilities on the lattice}
\label{sec:Compton}

Lattice QCD provides a way to study the polarisabilities of hadrons
from first principles. There are two ways to do this. The method most
reminiscent of the experimental situation is to study the (Euclidean
space) four point Green function defining the Compton scattering
tensor directly (the photon fields are amputated). By measuring the
large Euclidean time behaviour of this correlator, the hadron matrix
elements of the two vector currents can be extracted. In principle, by
calculating particular Lorentz components of the Compton tensor with
various different source and sink spin states, all six electromagnetic
and spin polarisabilities and their higher order and generalised
analogues can be extracted.  However, this is a complicated task,
requiring the evaluation a large number of quark propagator
contractions resulting from quark-line disconnected diagrams which are
statistically difficult to determine.  At present this approach is too
demanding for the available computational resources and has not been
attempted.

The second method is based on measuring the response of hadronic
states to fixed external fields. A number of exploratory quenched QCD
studies have been performed in this approach. The pioneering
calculations of
Refs.~\cite{Fucito:1982ff,Martinelli:1982cb,Bernard:1982yu,Aoki:1989rx,Aoki:1990ix,Shintani:2005du}
attempted to measure the nucleon axial couplings, magnetic moments and
electric dipole moments by measuring the linear shift in the hadron
energy as a function of an applied external weak or electromagnetic
field.  As discussed in the Introduction, various groups
\cite{Fiebig:1988en,Christensen:2004ca,Lee:2005dq,Zanottiprivatecommunication}
have also used this approach to extract electric and magnetic
polarisabilities in quenched QCD by measuring a quadratic shift in the
hadron energy in external electric and magnetic fields. The method is
not limited to electroweak external fields and can be used to extract
many matrix elements such as those that determine the moments of
parton distributions and the total quark contribution to the spin of
the proton \cite{Detmold:2004kw}. Here we focus on the electromagnetic
case.

The Euclidean space ($x_4 \equiv \tau$) effective action describing the gauge and parity
invariant interactions of a non-relativistic spin-half hadron of mass
$M$ and charge $q$ with a classical U(1) gauge field,
$A^\mu(\vec{x},\tau)$, is
\begin{eqnarray}
  \label{eq:eff_act}
  S_{\rm eff}[A]=\int d^3 x\,d\tau\, {\cal L}_{\rm eff}(\vec{x},\tau;A)\,,
\end{eqnarray}
for the Lagrangian
\begin{multline}
  \label{eq:eff_L}
  {\cal L}_{\rm eff}(\vec{x},\tau;A) = \Psi^\dagger(\vec{x},\tau)
  \Bigg[\left(\frac{\partial}{\partial \tau}+i\,q\,A_4\right)
+\frac{(-i \vec\nabla-\,q\,\vec{A})^2}{2M}  - \mu\, \vec{\sigma}\cdot\vec{H}
  +{2\pi}\left(\alpha\, \vec{E}^2 -\beta\, \vec{H}^2\right) 
\\
	-2\pi i \left( -\gamma_{E_1E_1} \vec{\sigma}\cdot\vec{E}\times\dot{\vec{E}}
    + \gamma_{M_1M_1} \vec{\sigma}\cdot\vec{H}\times\dot{\vec{H}}
    + \gamma_{M_1E_2} \sigma^i E^{ij}H^j
    + \gamma_{E_1M_2} \sigma^i H^{ij}E^j
\right)\Bigg] \Psi(\vec{x},\tau)  +\ldots \,,
\end{multline}
where $\vec{E}=- \frac{\partial}{\partial \tau} \vec{A}(\vec{x},\tau) -\vec{\nabla} A_4(\vec{x},\tau)$ and $\vec{H}= \vec{\nabla}\times \vec{A}(\vec{x},\tau)$
are the corresponding electric and magnetic fields, $\dot{X} = \frac{\partial}{\partial \tau} X$ denotes the Euclidean time derivative, 
$X^{ij}=\frac{1}{2}(\partial^i X^j+\partial^j X^i)$, and the ellipsis
denotes terms involving higher dimensional operators.%
By calculating
one- and two-photon processes with this effective Lagrangian, it is
clear that the constants that appear in Eq.~(\ref{eq:eff_L}) are
indeed the relevant magnetic moment and electromagnetic and multipole
polarisabilities \cite{Babusci:1998ww} [these are simply related to
the polarisabilities defined in the previous section as:
$\gamma_{E_1E_1}=-(\gamma_1+\gamma_3)$, $\gamma_{M_1M_1}=\gamma_4$,
$\gamma_{E_1M_2}=\gamma_3$ and $\gamma_{M_1E_2}=\gamma_2+\gamma_4$].
The Schr\"odinger equation correpsonding to Eq.~(\ref{eq:eff_L})
determines the energy of the particle in an external U(1) field in
terms of the charge, magnetic moment, and polarisabilities.  Higher
order terms in Eq.~(\ref{eq:eff_L}) (which contain in part the higher
order polarisabilities \cite{Holstein:1999uu}) can be neglected for
sufficiently weak external fields. For a magnetic field, the minimally
coupled terms generate towers of Landau levels and for a constant
electric field the same terms accelerate charged particles.

Lattice calculations of the energy of a hadron in an external U(1)
field are straight-forward. One measures the behaviour of the usual
two-point correlator on an ensemble of gauge configurations generated
in the presence of the external field. This changes the Boltzmann
weight used in selecting the field configurations from
$\det\left[\Dslash+m\right]\exp{\left[-S_g\right]}$ to
$\det\left[\Dslash +i\,\hat{Q}\,\slash\!\!\!\!A+m\right]\exp{\left[-S_g\right]}$,
where $\Dslash\,$ is the SU(3) gauge covariant derivative, $\hat{Q}$
is the quark electromagnetic charge operator, and $S_g$ is the usual
SU(3) gauge action.  Since calculations are required at a number of
different values of the field strength in order to correctly identify
shifts in energy from the external field, this is a relatively demanding
computational task (although it is at least conceptually simpler than
studying the four-point function). In general one must worry about the
positivity of the fermionic determinant calculated in the presence of
a background field, however for weak fields, positivity is preserved.
The exploratory studies of
Refs.~\cite{Fucito:1982ff,Martinelli:1982cb,Bernard:1982yu,Fiebig:1988en,Aoki:1989rx,Aoki:1990ix,Shintani:2005du,Christensen:2004ca,Lee:2005dq,Zanottiprivatecommunication}
used quenched QCD in which the gluon configurations do not feel the
presence of the U(1) field as the quark determinant is absent. In this
case, the external field can be applied after the gauge configurations
had been generated and is simply implemented by multiplying the SU(3)
gauge links of each configuration by link variables corresponding to
the fixed external field: $\{U^{\mu}_{\alpha}(x)\}\longrightarrow
\{U^{\mu} _{\alpha}(x) \exp[i\, e\,a\, A^\mu]\}$, where $a$ is the
lattice spacing. These studies are interesting in that they provide a
proof of the method, however the values of the polarisabilities
extracted have no connection to those measured in experiment.

It is clear from Eq.~(\ref{eq:eff_L}) that all six polarisabilities
can be extracted using suitable space and time varying background
fields if the shift of the hadron energy at second order in the
strength of the field can be determined. One can also see this because
the Compton tensor appears explicitly as the second-order connected
term in the expansion of hadronic two-point correlation function in
weak background fields \cite{Detmold:2004kw}. Previous studies
\cite{Fiebig:1988en,Christensen:2004ca,Lee:2005dq,Zanottiprivatecommunication}
have employed constant electric and magnetic fields to determine the
corresponding polarisabilities in quenched QCD. Here we perform a more
general analysis to show how the spin polarisabilities and the
electric polarisabilities of charged particles can be obtained.

In order to determine the polarisabilities, we consider lattice
calculations of the two-point correlation function
\begin{eqnarray}
  \label{eq:correlator}
  C_{s s^\prime}(\vec{p},\tau;A)=\int d^3x\, e^{i\vec{p}\cdot\vec{x}}
\langle 0| \chi_s(\vec{x},\tau)\chi^\dagger_{s^\prime}(0,0) |0\rangle_A\,,
\end{eqnarray}
where $\chi_s(\vec{x},\tau)$ is an interpolating field with the quantum
numbers of the hadron under consideration (we will focus on the
nucleons) with $z$ component of spin, $s$, and the correlator is
evaluated on the ensemble of gauge configurations generated with the
external field, $A^\mu$.

For uncharged hadrons at rest in constant electric and magnetic
fields, it is simple to show that this correlator falls off
exponentially at large times with an energy given by the appropriate
terms in Eq.~(\ref{eq:eff_L}) owing to the constancy of the effective
Hamiltonian.  However for space-time varying fields, charged particles
or states of non-zero $\vec{p}$, a more general analysis is needed.
This is most easily formulated using the effective field theory (EFT)
defined by Eq.~(\ref{eq:eff_L}).  For weak external fields (such that
higher order terms in Eq.~(\ref{eq:eff_L}) can be safely neglected),
the small $\vec{p}$ and large $\tau$ dependence of this QCD correlation
function is reproduced by the equivalent correlator calculated in the
effective theory corresponding to the Lagrangian,
Eq.~(\ref{eq:eff_L}). That is
\begin{eqnarray}
  \label{eq:eft_correlator}
  C_{s s^\prime}(\vec{p},\tau;A)&=&\int d^3x\, e^{i\vec{p}\cdot\vec{x}}
\frac{1}{{\cal Z}_{\rm eff}[A]}\int {\cal D}\Psi^\dagger {\cal D}\Psi
\,\Psi_s(\vec{x},\tau)\Psi^\dagger_{s^\prime}(0,0)
\exp\left(-S_{\rm
    eff}[A]\right)\,,
\end{eqnarray}
where ${\cal Z}_{\rm eff}[A]=\int {\cal D}\Psi^\dagger {\cal D}\Psi
\exp\left({-{S}_{\rm eff}[A]}\right)$.  Since the right-hand side of
Eq.~(\ref{eq:eft_correlator}) is completely determined in terms of the
charge, magnetic moment and polarisabilities that we seek to extract,
fitting lattice calculations of $C_{s s^\prime}(\vec{p},\tau;A)$ in a
given external field to the effective field theory expression will
enable us to determine the appropriate polarisabilities.  In the above
equation we have assumed that the ground state hadron dominates the
correlator at the relevant times. For weak fields this will be the
case. However one can consider additional terms in the effective
Lagrangian that describe the low excitations of the hadron spectrum
that have the same quantum numbers as the hadron under study. This
will lead to additive terms in Eq.~(\ref{eq:eft_correlator}) that
depend on the mass, magnetic moment and polarisabilities of the
excited hadron instead of those of the ground state. With precise
lattice data, the properties of these excited states can also be
determined.

In many simple cases such as constant or plane-wave external fields,
the EFT version of $C_{s s^\prime}(\vec{p},\tau;A)$ can be determined
analytically in the infinite volume, continuum limit
\cite{Schwinger:1951nm}. However in finite lattice spacing and at
finite volume, calculating $C_{s s^\prime}(\vec{p},\tau;A)$ in the EFT
becomes more complicated. In order to determine the EFT correlator, we
must invert the matrix ${\cal K}$ defined by
\begin{eqnarray}
  \label{eq:S_latt}
  S_{\rm
  latt}[A]=\sum_{\vec{x},\tau_x}\sum_{\vec{y},\tau_y}\sum_{s,s^\prime}\Psi^\dagger_s(\vec{x},\tau_x)
{\cal K}_{ss^\prime}[\vec{x},\tau_x,\vec{y},\tau_y;A] \Psi_s(\vec{y},\tau_y)\,,
\end{eqnarray}
where $S_{\rm latt}[A]$ is a discretisation of the EFT action in which
derivatives are replaced by finite differences (the time derivative is
given by a forward difference as we can ignore anti-particles). ${\cal
  K}$ has dimension $4N_{l}^2$ where $N_l$ is the number of lattice
sites.  For the most general space-time varying external field, this
must be inverted numerically; given a set of lattice results for the
correlator, Eq.~(\ref{eq:eft_correlator}) is repeatedly evaluated for
varying values of the polarisabilities until a good description of the
lattice data is obtained.

\begin{figure}[!t]
  \centering
  \includegraphics[width=0.8\columnwidth]{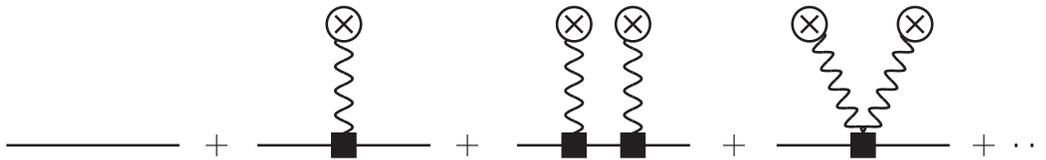}
  \caption{Perturbative expansion of the hadron propagator in an
    external field. }
  \label{fig:EFT}
\end{figure}

For weak fields such that $|A^\mu(\vec{x},\tau)|^2\ll\L_{\rm QCD}^2$ for
all $\vec{x}$ and $\tau$, a perturbative expansion of ${\cal K}^{-1}$ in
powers of the field can be used. This corresponds to the series of
diagrams in Fig.~\ref{fig:EFT}. 
To extract all six polarisabilities using such an analysis, we need to
consider a number of different fields; lattice calculations of the
correlators in Eq.~(\ref{eq:correlator}) using
\begin{eqnarray}
  \label{eq:fields}
  A^\mu_{(1)}(x) =\begin{pmatrix} i a_1 \tau \\0\\0\\0\end{pmatrix},\quad
  A^\mu_{(2)}(x) =\begin{pmatrix}-\frac{a_2}{2} x_2\\ \frac{a_2}{2} x_1\\ 0
    \\0\end{pmatrix},\quad
  A^\mu_{(3)}(x) =\begin{pmatrix}0 \\ -  \frac{i}{a} a_3 \tau \,x_3\\ -b_3 x_1\\0
    \end{pmatrix}, 
\\ \nonumber
  A^\mu_{(4)}(x) =\begin{pmatrix} - \frac{i}{a}a_4 \tau \,x_2 \\ 0\\
    -\frac{1}{2}\,b_4 x_1 \\0\end{pmatrix},\quad 
  A^\mu_{(5)}(x) =\begin{pmatrix} \frac{1}{a}a_5\, x_2\, x_1 \\
    \frac{1}{2a}a_5\, x_2^2 \\ i b_5 \tau
    \\0\end{pmatrix},\quad 
  A^\mu_{(6)}(x) =\begin{pmatrix} - \frac{1}{2a}a_6 \tau^2 \\ -  i\frac{1}{2}
    b_6 \tau \\ 0 \\0\end{pmatrix}\,,
\end{eqnarray}
for a number of different choices for the strength parameters, $a_i$
and $b_i$ (with $|a_i|,\,|b_i|\ll\L_{\rm QCD}^2$), are sufficient to
determine the full set of polarisabilities.%
\footnote{These fields correspond to real $\vec{E}$ and $\vec{H}$ fields in Minkowski-space for real-valued $a_i$ and $b_i$.  Since periodic spatial
  boundary conditions are envisaged for the link variables, there are
  quantisation conditions that must be satisfied by the $a_i$ 
  \cite{Martinelli:1982cb,Bernard:1982yu}. For example, $q_i
  a_2=\frac{2\pi\, n}{a\,L}$ for each of the quark charges $q_i$. 
  The more complicated fields in Eq.~(\ref{eq:fields}) require two
  parameters to satisfy these conditions.  
}  
By measuring correlators for different spin
configurations (including those that flip spin), we can reduce the
number of fields required to extract the polarisabilities.

As an example, the behaviour of the correlator in the field
$A^\mu_{(1)}(x)$ (which corresponds to a constant electric field in
the $x_1$ direction) is given by
\begin{eqnarray}
  C_{s s^\prime}(\vec{p},\tau;A_{(1)}) &=&
  \delta_{s,s^\prime}\exp\left\{
- \frac{a_1\,\tau}{6M}\left[ a_1
  \left(q^2 \tau^2 + 12 M \pi  \alpha \right) - 3 i q\, \tau\,p_1\right]\right\}
e^{-M\,\tau} e^{-\frac{\tau}{2M}|\vec{p}|^2} +{\cal O}(a_1^3)
\notag
\\
&\stackrel{|\vec{p}|\to0}{\longrightarrow}&
\delta_{s,s^\prime}\exp\left[-(M+2\pi\alpha
  a_1^2)\tau-\frac{q^2a_1^2}{6M}\tau^3\right] +{\cal O}(a_1^3) \,.
  \label{eq:A1field}
\end{eqnarray}
In this case, the perturbative series has been resummed exactly in the
continuum, infinite volume limit and the higher order corrections come
from terms omitted in Eq.~(\ref{eq:eff_L}). For electrically neutral
particles, the exponential fall-off of this correlator determines the
polarisability $\alpha$ once the mass $M$ has been measured in the
zero-field case.  When a charged particle is placed in such a field it
undergoes continuous acceleration in the $x_1$ direction (this is
described by the $\tau^3$ term in the exponent).  However at times small
compared to $\frac{\sqrt{6M}}{q\, a_1}$, the correlator essentially
falls off exponentially. Matching the behaviour of
Eq.~(\ref{eq:A1field}) to lattice data for a charged hadron will again
enable us to determine the electric polarisability, $\alpha$.

As a second analytic example, we consider one of the multipole
polarisabilities. In the presence of the field $A_{(6)}^\mu(x)$, which
corresponds to a more complicated electric field
$\vec{E}_{(6)}(x)=(\frac{a_6}{a} \tau, i\frac{b_6}{2},0)$, we find that
\begin{eqnarray}
  \label{eq:A6exp}
  \frac{C_{\uparrow\uparrow}(\vec{p},\tau;A_{(6)})}
  {C_{\downarrow\downarrow}(\vec{p},\tau;A_{(6)})}
  &=& \exp\left[\frac{2\pi}{a}\,a_6\,b_6\,\gamma_{E_1E_1}\,  \tau \right]+\ldots \,,
\end{eqnarray}
independent of $\vec{p}$ and the ellipsis denotes terms cubic in the
field that have been neglected in Eq.~(\ref{eq:eff_L}).  Whilst the
individual correlators, $C_{\uparrow\uparrow}$ and
$C_{\downarrow\downarrow}$, have relatively complicated time-momentum
behaviour involving $q$ and $\alpha$ as well as $\gamma_{E_1E_1}$,
this becomes very simple in the ratio and $\gamma_{E_1E_1}$ can be
determined cleanly.

Analogous results can be derived for the other fields in
Eq.~(\ref{eq:fields}), however to take into account the finite lattice
spacing and periodic finite volume nature of the underlying lattice
simulations to which the EFT description is matched, the correlator is
most easily calculated by inverting the matrix ${\cal K}$ numerically.
This also allows for more general choices of fields. If we seek to
extract higher order polarisabilities, the Lagrangian in
Eq.~(\ref{eq:eff_L}) must be extended to include higher dimension
operators \cite{Holstein:1999uu}.  At this order, relativistic
corrections and three-photon couplings also need to be included.
Correlation functions similar to those in Eq.~(\ref{eq:correlator})
involving two different external momenta will allow us to also extract
the generalised polarisabilities \cite{Guichon:1995pu}.

%
%
%
%
%
%
%
%
%
%
%
%
\section{Heavy baryon \xpt\ }
\label{sec:hbxpt}

To calculate the quark mass and volume dependence of the nucleon
polarisabilities, we use heavy baryon chiral perturbation theory
(HB$\chi$PT) as was first constructed in
Refs.~\cite{Jenkins:1990jv,Jenkins:1991ne,Jenkins:1991es,Bernard:1993nj}.
In current lattice calculations, valence and sea quarks are often
treated differently, with sea quarks either absent (quenched QCD) or
having different masses than the valence quarks (partially-quenched
QCD).\footnote{At finite lattice spacing, different actions can even
  be used for the different quark sectors (e.g., staggered sea quarks
  and domain wall valence quarks).  As was shown in
  Refs.~\cite{Beane:2003xv,Arndt:2003vd,Tiburzi:2005is}, the lattice
  spacing corrections to baryon electromagnetic properties are
  expected to be small, as they can not enter at tree level, and for
  current simulations with $a \L_{QCD}^2 \sim m_q$, they generally
  enter at leading loop order through valence-sea meson masses.  In
  our work we assume a continuum extrapolation has been performed.}
The extensions of \hbxpt\ to quenched \hbxpt\ 
\cite{Labrenz:1996jy,Savage:2001dy} and partially quenched \hbxpt\ 
\cite{Chen:2001yi,Beane:2002vq} to accommodate these modifications are
also well established and have been used to calculate many baryon
properties.  In this section, we will primarily focus on the two
flavour partially-quenched theory and briefly introduce the relevant
details following the conventions set out in Ref.~\cite{Beane:2002vq}.
Since QCD is a special limit of the partially-quenched theory, our
discussion also encompasses two flavour \xpt.  Additional
complications in quenched \xpt\ are relegated to Appendix~\ref{A4}.

%
%
%
%
%
%
%
%
%
%
%
%
\subsection{Pseudo-Goldstone mesons}

We consider a partially-quenched theory of valence ($u$, $d$), sea
($j$, $l$) and ghost ($\tilde u,\,\tilde d$) quarks with masses
contained in the matrix
\begin{equation}
\label{eq:Mq_def}
m_Q = {\rm  diag}(m_u,m_d,m_j,m_l,m_{\tilde u},m_{\tilde d})\,,
\end{equation}
where $m_{\tilde u,\tilde d}=m_{u,d}$ such that the path-integral
determinants arising from the valence and ghost quark sectors exactly
cancel. The corresponding low-energy meson dynamics are described by
the \pqxpt\ Lagrangian. At leading order this is given by
\begin{eqnarray}
{\cal L }_\Phi & = & 
{f^2\over 8} 
{\rm str}\left[ {\cal D}^\mu\Sigma^\dagger {\cal D}_\mu\Sigma \right] + 
\lambda {f^2\over 4} 
{\rm str}\left[ m_Q\Sigma^\dagger + m_Q^\dagger\Sigma \right]
 + 
\alpha_\Phi {\cal D}^\mu\Phi_0 {\cal D}_\mu\Phi_0 - m_0^2\Phi_0^2
\, ,
\label{eq:PGBlagrangian}
\end{eqnarray}
where the pseudo-Goldstone mesons are embedded non-linearly in
\begin{equation}
\label{eq:Sigma_def}
\Sigma = \xi^2= \exp{\left(\frac{2\,i\,\Phi}{f}\right)} ,
\end{equation}
with the matrix $\Phi$ given by
\begin{equation}
  \label{eq:Phi_def}
\Phi = \begin{pmatrix} M & \chi^\dagger \cr \chi &\tilde{M} \end{pmatrix}
\,,
\end{equation}
and
\begin{eqnarray}
  \label{eq:Mchi_def}
        M =\begin{pmatrix}
                \eta_u & \pi^+ & \phi_{uj} & \phi_{ul} \\
                \pi^- & \eta_d & \phi_{dj} & \phi_{dl} \\
                \phi_{ju} & \phi_{jd} & \eta_j & \phi_{jl} \\
                \phi_{lu} & \phi_{ld} & \phi_{lj} & \eta_l
                \end{pmatrix}\ ,
&  \hspace*{15mm}
\tilde{M}=\begin{pmatrix} \tilde\eta_u    & \tilde\pi^+        \\
                          \tilde\pi^-     & \tilde\eta_d       
                          \end{pmatrix}\,, 
\hspace*{15mm} &
        \chi=\begin{pmatrix} 
                \phi_{\tilde{u}u} & \phi_{\tilde{u}d} & \phi_{\tilde{u}j} & \phi_{\tilde{u}l} \\
                \phi_{\tilde{d}u} & \phi_{\tilde{d}d} & \phi_{\tilde{d}j} & \phi_{\tilde{d}l} 
                \end{pmatrix}\,,
\end{eqnarray}
and where $\Phi_0=\str\left(\Phi\right)/\sqrt{2}$.  The upper left
$2\times2$ block of $M$ corresponds to the usual valence--valence
mesons, the lower right to sea--sea mesons and the remaining entries
of $M$ to valence--sea mesons. Mesons in $\tilde{M}$ are composed of
ghost quarks and ghost anti-quarks and thus bosonic.  Mesons in $\chi$
contain ghost--valence or ghost--sea quark--anti-quark pairs and are
fermionic. In terms of the quark masses, the tree-level meson masses
are given by
\begin{eqnarray}
  \label{eq:mesonmass_def}
  m^2_{\Phi_{ij}}=m_{{ Q}_i { Q}_j}^2 = 
\lambda\left[ \left(m_Q\right)_{ii}
    + \left(m_Q\right)_{jj} \right] \,,
\end{eqnarray}
where ${ Q}=(u,\,d,\,j,\,l,\,\tilde{u},\,\tilde{d})$.
The terms proportional to $\alpha_\Phi$ and $m_0$ in
Eq.~(\ref{eq:PGBlagrangian}) involve the flavour singlet field and are
only relevant in the quenched theory (see Appendix~\ref{A4}); in both
\pqxpt\ and \xpt\, the singlet meson acquires a large mass through the
strong U(1)$_A$ anomaly and can be integrated out, leading to a
modified flavour neutral propagator that contains both single and
double pole structures~\cite{Sharpe:2000bn}.

In the above Lagrangian, we have minimally coupled electromagnetism
(the U(1) gauge field is again denoted by $A^\mu$ and its field
strength tensor $F^{\mu\nu}=\partial^\mu A^\nu-\partial^\nu A^\mu$) to
the theory through the chiral, and U(1) gauge covariant derivative
\begin{equation}
\label{eq:cov_deriv_def}
  {\cal D}^{\mu} = \partial^{\mu}
  + \left [ {\cal V}^{\mu},\quad \right ] \,,
\end{equation}
with the vector current
\begin{eqnarray}
  \label{eq:vector_current}
  {\cal V}^{\mu} &=& {1\over 2}\left[\ \xi\left(\partial^\mu-i e {\cal
        Q} A^\mu\right)\xi^\dagger \ + \  
\xi^\dagger\left(\partial^\mu-i e {\cal Q} A^\mu\right)\xi \ \right],
\end{eqnarray}
depending on the quark charge matrix, ${\cal Q}$. In coupling
electromagnetism to this theory, we must specify how the quark charges
are extended to the partially-quenched theory. We choose:
\begin{eqnarray}
  \label{eq:7}
  {\cal Q} &=& {\rm diag}(q_u,q_d,q_j,q_l,q_u,q_d)\,,
\end{eqnarray}
though other arrangements are possible.  However, one must set
$q_j+q_l\ne0$ in order to retain sensitivity to the full set of LECs
that appear in two flavour \xpt~\cite{Tiburzi:2004mv,Detmold:2005pt}.
In addition to the Lagrangian, Eq.~(\ref{eq:PGBlagrangian}), the
anomalous couplings of the Wess-Zumino-Witten Lagrangian
\cite{Wess:1971yu,Witten:1983tw} will also contribute to the spin
polarisabilities. These terms are described below.

%
%
%
%
%
%
%
%
%
%
%
%
\subsection{Baryons}

In SU(4$|$2) \hbxpt, the physical nucleons (those composed of three
valence quarks) are embedded in a {\bf 70}-dimensional representation
of the flavour group described by a three index flavour-tensor, ${\cal
  B}$ \cite{Beane:2002vq,Chen:2001yi}.  Since the mass-splitting
between the nucleon and $\Delta$-isobar,
$\Delta=M_\Delta-M_N\sim300$~MeV, is comparable to the physical pion
mass (and less than pion masses used in current lattice simulations),
the $\Delta$-isobar must be included in the theory. These fields are
represented in a totally symmetric three index flavour-tensor ${\cal
  T}^\mu$ (a Rarita-Schwinger field) transforming as a {\bf
  44}-dimensional representation of SU(4$|$2).  The mass-splitting
$\Delta$ is small compared to the chiral symmetry breaking scale, and
in this work we treat $\D \sim m_\pi$ in the power counting
\cite{Jenkins:1991ne}.  For additional details, see
Refs.~\cite{Labrenz:1996jy,Beane:2002vq,Chen:2001yi}.

The relevant part of leading-order Lagrangian describing these baryons
and their interactions with Goldstone mesons is
\begin{eqnarray}
{\cal L}_B^{(0)} & = & 
i\left(\overline{\cal B} v\cdot {\cal D} {\cal B}\right)
 - i \left(\overline{\cal T}^\mu v\cdot {\cal D} {\cal T}_\mu\right)
+ \Delta\ \left(\overline{\cal T}^\mu {\cal T}_\mu\right)
\nonumber\\
 && + 2\alpha \left(\overline{\cal B} S^\mu {\cal B} {\cal A}_\mu\right)
 +  2\beta \left(\overline{\cal B} S^\mu {\cal A}_\mu {\cal B} \right)
 +  2{\cal H} \left(\overline{\cal T}^\nu S^\mu {\cal A}_\mu {\cal T}_\nu \right)
+ \sqrt{3\over 2}{\cal C} 
\left[
\left( \overline{\cal T}^\nu {\cal A}_\nu {\cal B}\right) + 
\left(\overline{\cal B} {\cal A}_\nu {\cal T}^\nu\right) \right]
\,,
\label{eq:free_lagrangian}
\end{eqnarray}
where $v^\mu$ is the baryon velocity, $S^\mu$ is the covariant
spin-vector \cite{Jenkins:1990jv,Jenkins:1991es} and ${\cal D}^\mu$ is
the chiral, and U(1) covariant derivative~\cite{Labrenz:1996jy}.  The
axial-vector current is given by
\begin{eqnarray}
{\cal A}^\mu \ =\  {i\over 2}\left[\ \xi\left(\partial^\mu-i e {\cal
        Q} A^\mu\right)\xi^\dagger \ - \  
\xi^\dagger\left(\partial^\mu-i e {\cal
        Q} A^\mu\right)\xi \ \right]
\, .
\label{eq:Amu}
\end{eqnarray}
The various flavour contractions (indicated by the parentheses) are
defined as in Ref.~\cite{Beane:2002vq}.  In order for ${\cal T}^\mu$
to correctly describe the spin-$3/2$ sector, the constraints $v\cdot
{\cal T}= S \cdot {\cal T}=0$ are used.  The partially quenched
Lagrangian, Eq.~\eqref{eq:free_lagrangian}, contains one more operator
than the corresponding two flavor \CPT\ Lagrangian.  To determine the
relation of the partially quenched operators to those of SU(2) \xpt,
one simply restricts the flavor indices of all the operators in
Eq.~\eqref{eq:free_lagrangian} to the valence sector, leading to
\begin{eqnarray}
        \a = \frac{4}{3}g_A + \frac{1}{3}g_1\,,\quad && \quad
                \b = \frac{2}{3}g_1 -\frac{1}{3}g_A , \nonumber\\
        \mathcal{C} = -g_{\Delta N}\,, \quad && \quad 
                \mathcal{H} = g_{\D\D} .
\end{eqnarray}
Here, $g_1$ is the coupling of the nucleons to the SU(2) singlet-meson
field, which decouples in \CPT.  Thus when the QCD limit of the
partially quenched theory is taken, where $m_{j}=m_{u}$ and
$m_{l}=m_{d}$, all dependence on $g_1$ vanishes (in the quenched
theory, quantities may still depend on this parameter, see
Appendix~\ref{A4}).

As with the mesons, at leading order the photon is minimally coupled
to the baryons with fixed coefficients.  At the next order in the
expansion there are a number of new electromagnetic gauge invariant
operators which contribute to the Compton amplitude and the
polarisabilities.  Here, we display the relevant terms at this order,
\begin{eqnarray}
        {\cal L}_{B}^{(1)}& = &
                \frac{i\,e}{ 2 M_N} F_{\mu\nu}
                \left[
                        \mu_\alpha \left( \overline{\cal B} \left[S^{\mu},S^{\nu}\right]  {\cal B}  
                                {\cal Q}_{\xi+} \right) 
                        + \mu_\beta \left( \overline{\cal B}
                          \left[S^{\mu},S^{\nu}\right]
                                {\cal Q}_{\xi+} {\cal B} \right)
                        +  \mu_\gamma  {\rm str}\left[ {\cal Q}_{\xi+} \right]
                 \left( \overline{\cal B} \left[S^{\mu},S^{\nu}\right] {\cal B} \right)
                 \right] \nonumber \\
                &&+ \sqrt{\frac{3}{2}} \mu_T \frac{i e}{ 2 M_N} F_{\mu\nu}
                \left[
                        \left( \overline{\cal B} S^\mu {\cal Q}_{\xi+} {\cal T}^\nu \right) 
                +\left( \overline{\cal T} {}^\mu S^\nu  {\cal Q}_{\xi+} {\cal B} \right) 
                \right],
\label{L1}
\end{eqnarray}
where $\mu_{\alpha,\beta,\gamma}$ are magnetic moment coefficients
\cite{Savage:2001dy,Beane:2002vq}, $\mu_T$ is the coefficient of the
M1 transition {\bf 70}--{\bf 44} operator
\cite{Arndt:2003we,Arndt:2003vd} and
\begin{eqnarray}
  \label{eq:1}
  {\cal Q}_{\xi^\pm} &=& {1\over 2} 
\left( \xi^\dagger {\cal Q}\xi \pm \xi {\cal Q}\xi^\dagger
\right).
\end{eqnarray}
The partially quenched magnetic moment coefficients are related to the
isoscalar and isovector magnetic coefficients, $\mu_0$ and $\mu_1$, in
standard two flavour \CPT\ as
\begin{equation}
        \mu_0 = \frac{1}{6} \Big( \mu_\a +\mu_\b +2\mu_\g \Big), \quad\quad
        \mu_1= \frac{1}{6} \Big( 2\mu_\a -\mu_\b \Big),
\end{equation}
where the \CPT\ Lagrangian describing the magnetic moments of the
nucleons [the proton and neutron magnetic moments are
$\mu_{p,n}=\frac{1}{2}(\mu_0\pm \mu_1)$] is given by
\begin{eqnarray}
        {\cal L} &=& \frac{i e}{2 M_N} F_{\mu\nu} \left( \mu_0
          \overline{N} \left[S^{\mu},S^{\nu}\right] N 
                                +\mu_1 \overline{N} \left[S^{\mu},S^{\nu}\right]
                                \t^3_{\xi+} N \right),
\end{eqnarray}
for $\t^a_{\xi\pm} = \frac{1}{2} \left( \xi^\dagger \t^a \xi \pm \xi
  \t^a \xi^\dagger \right)$.  

There are other operators formally at this order which do not
contribute to the polarisabilities at the order to which we work.
There are kinetic operators and higher dimensional couplings of the
baryons to the axial current whose coefficients are exactly fixed by
the reparameterisation invariance of the baryon
four-momentum~\cite{Luke:1992cs,Tiburzi:2005na}. These operators give
the $Z$ dependent pieces of the Compton amplitudes in
Eq.~(\ref{eq:amplitudes}). There are also additional operators with
unconstrained coefficients such as $ \left( \bar{\cal B}\, {\cal A}
  \cdot {\cal A}\, {\cal B} \right)$ that contribute to the Compton
amplitude at higher order.  In two-flavor \CPT\ there are two such
operators, and in the SU(4$|$2) case there are ten
\cite{Tiburzi:2005na}.

The leading operators which contribute to the electromagnetic
polarisabilities at tree level occur at ${\cal O}(Q^4)$ and are given
by the general form,
\begin{equation*}
        \frac{e^2F_{\mu\rho}F_{\nu}^{\;\rho}}{\L_\c^3}
                \left( \overline{\cal B}\, \Gamma^{\mu\nu} {\cal
                    Q}_{\xi \pm}^2 {\cal B} \right), 
\end{equation*}
(where the $\Gamma^{\mu\nu}$ are spin structures) while the leading
tree-level contributions to the spin polarisabilities occur at ${\cal
  O}(Q^5)$. The complete set of such operators in the case of
two-flavour \CPT\ is given in Ref.~\cite{Fettes:2000gb}. Again there
are significantly more such operators in \qxpt\ and \pqxpt. We do not
explicitly show these operators, as they do not contribute at the
order we are working and will not modify volume dependence until
${\cal O}(Q^6)$.

%
%
%
%
%
%
%
%
%
%
%
%
\section{Nucleon polarisabilities}
\label{sec:pols}

\begin{figure}[!t]
\centering
\begin{tabular}{cc}
        \includegraphics[width=0.35\columnwidth]{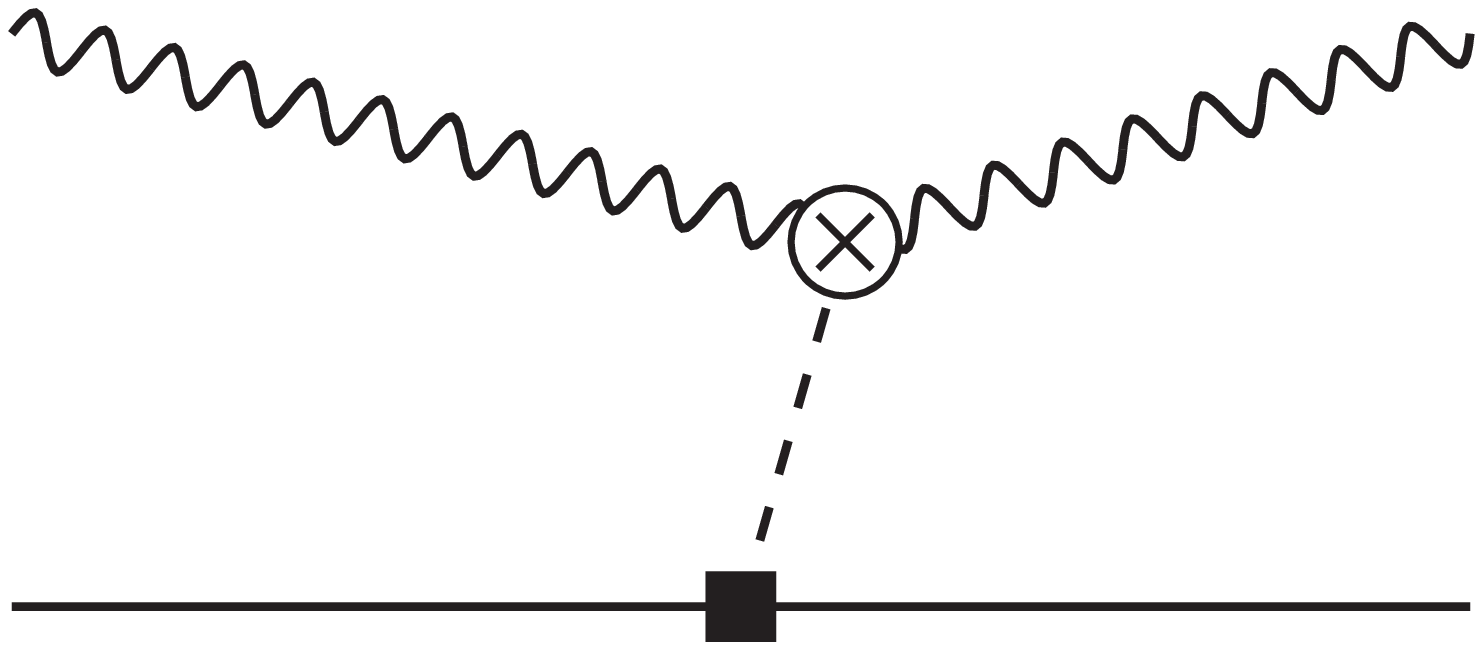}
        &\includegraphics[width=0.35\columnwidth]{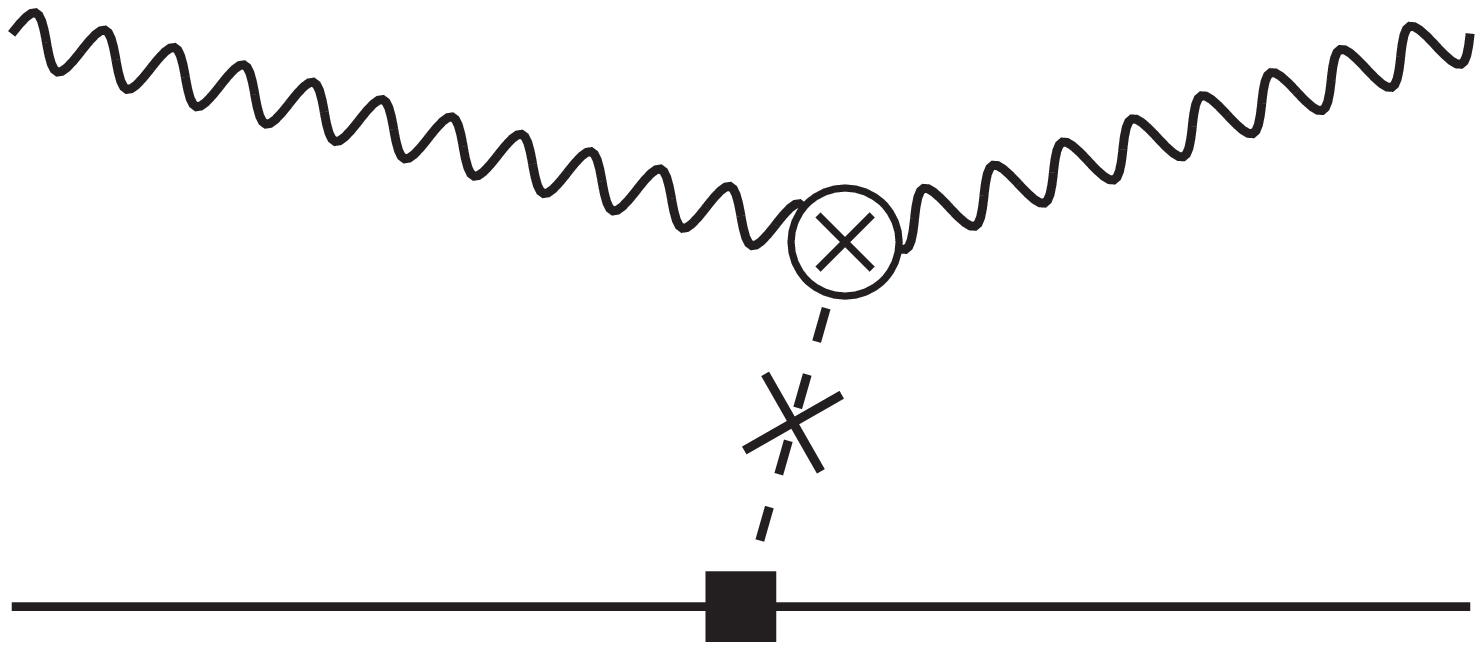}
\end{tabular}
\caption{Anomalous contributions to the polarisabilities. The crossed
  circle corresponds to the insertion of an operator from the
  Wess-Zumino-Witten Lagrangian, Eq.~(\ref{eq:LWZWPQ}), and the
  crossed meson line corresponds to a hairpin interaction
  \protect\cite{Sharpe:1992ft}.}
\label{fig:anom}
\end{figure}

\begin{figure}[!t]
\centering
\begin{tabular}{cc}
        \includegraphics[width=0.35\columnwidth]{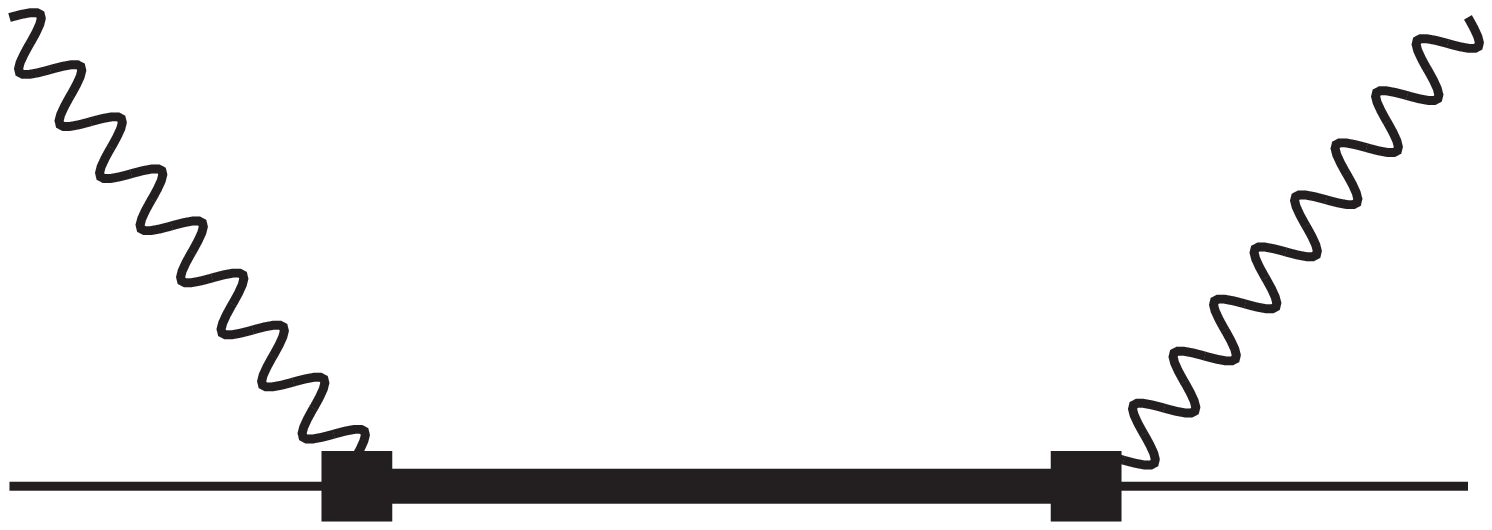} 
        &\includegraphics[width=0.35\columnwidth]{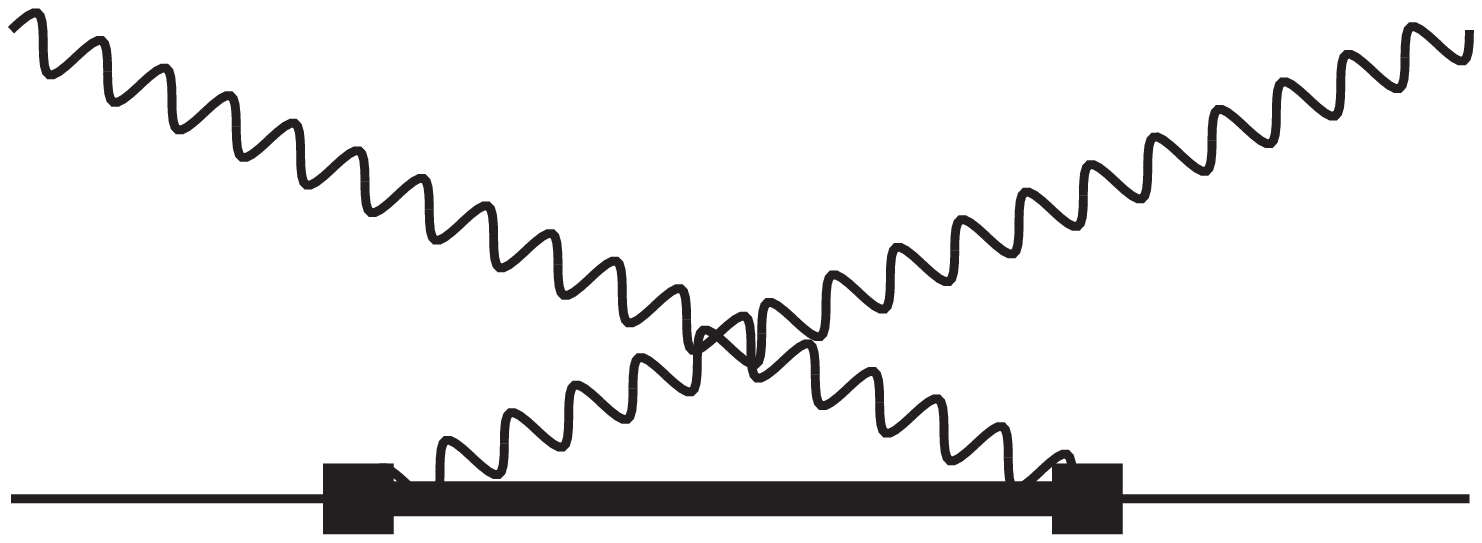} \\
\end{tabular}
\caption{Born diagrams involving internal {\bf 44}-plet states that
  give contribution to the polarisabilities.}
\label{fig:Born}
\end{figure}

\begin{figure}[!t]
  \centering
\begin{tabular}{ccc}
        \includegraphics[width=0.3\columnwidth]{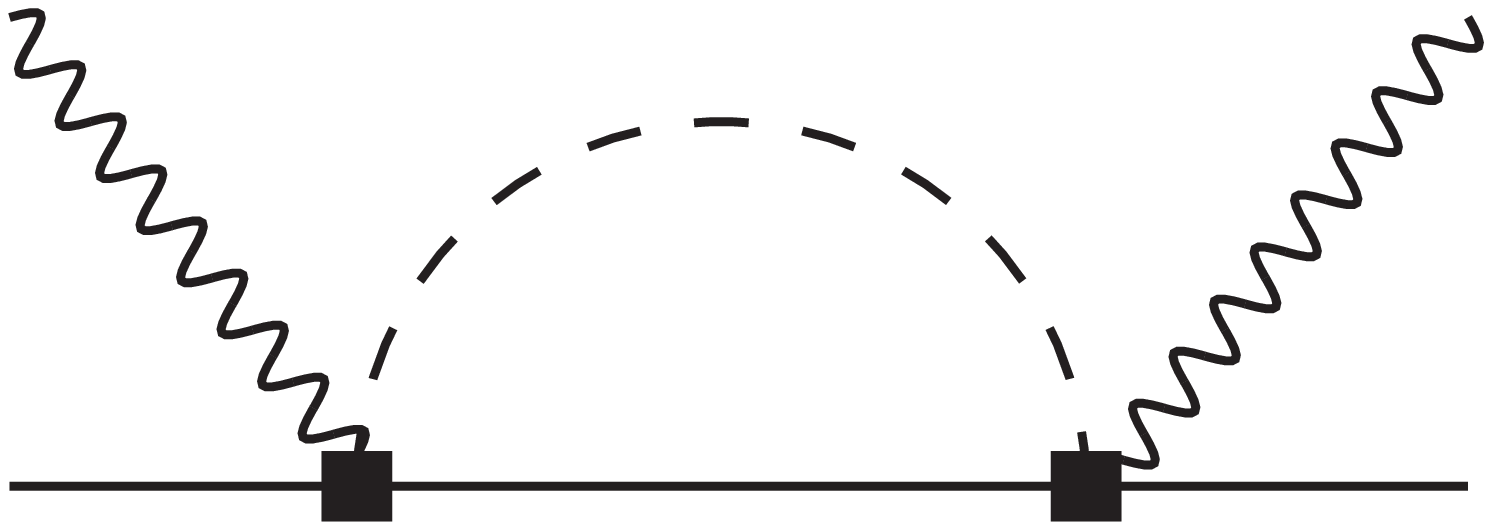}
        & \includegraphics[width=0.3\columnwidth]{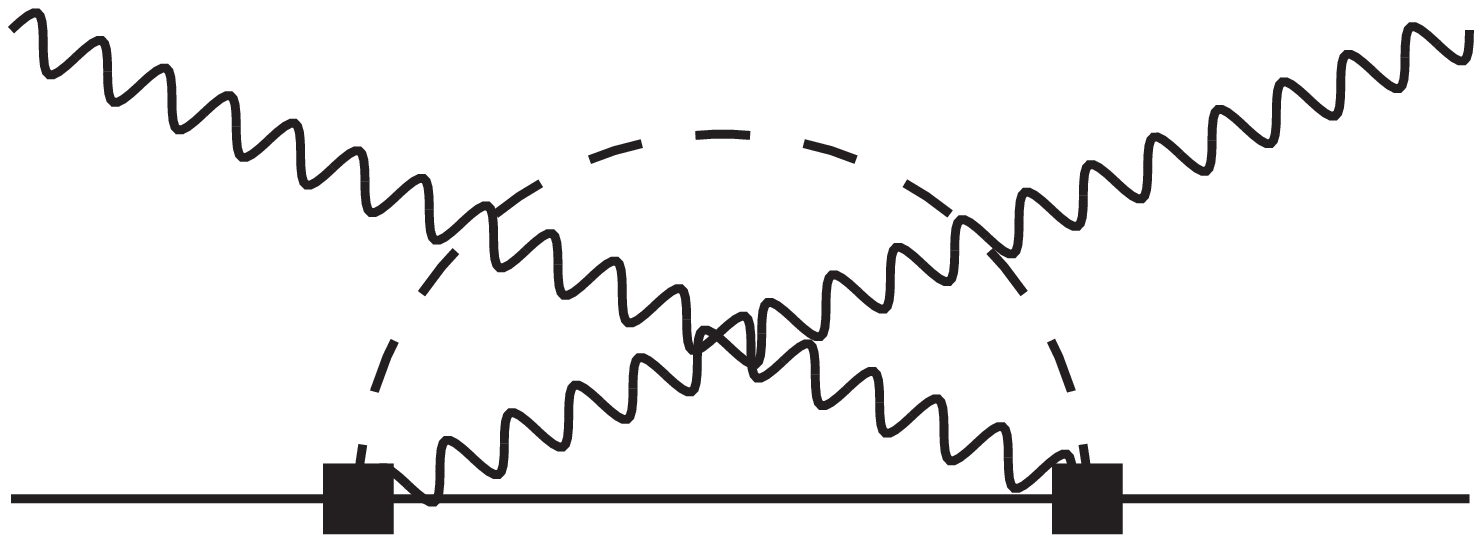} 
        & \includegraphics[width=0.3\columnwidth]{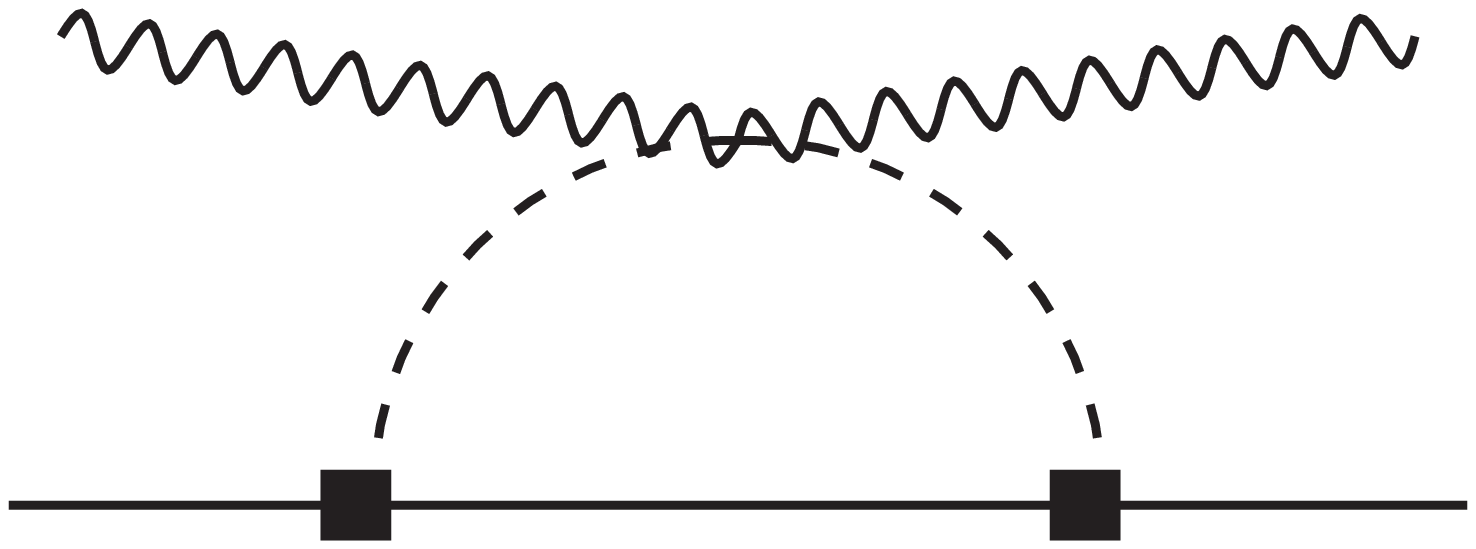} \\
        (a) & (b) &(c) \\ \\
        \includegraphics[width=0.3\columnwidth]{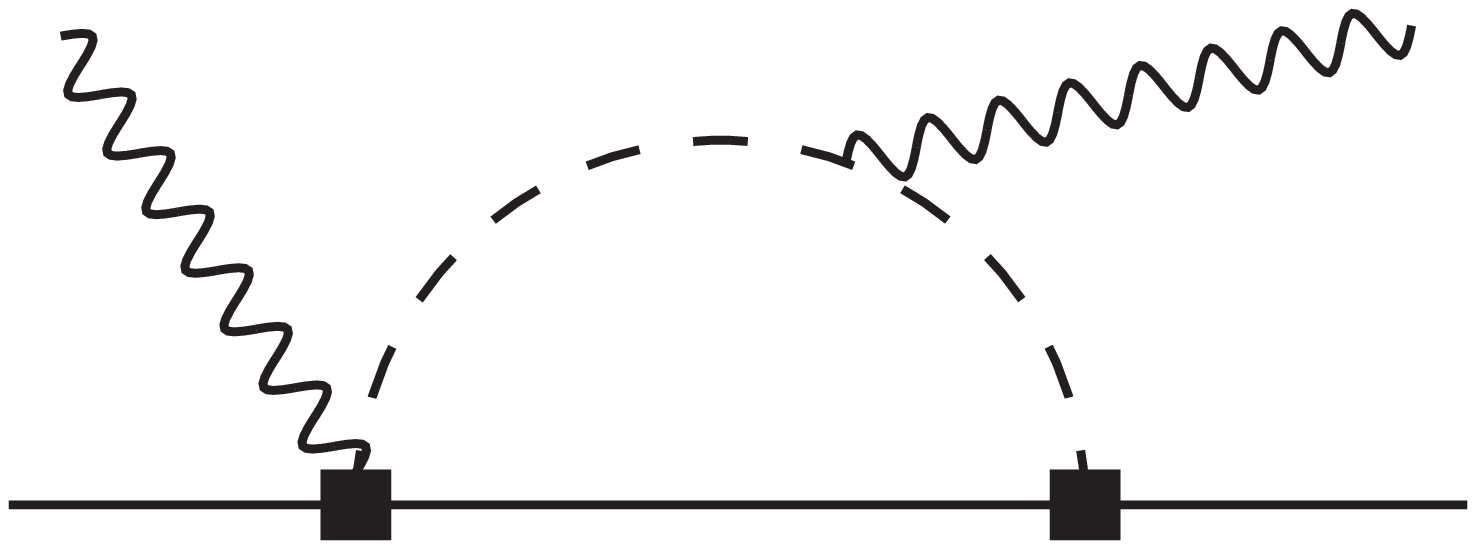} 
        &\includegraphics[width=0.3\columnwidth]{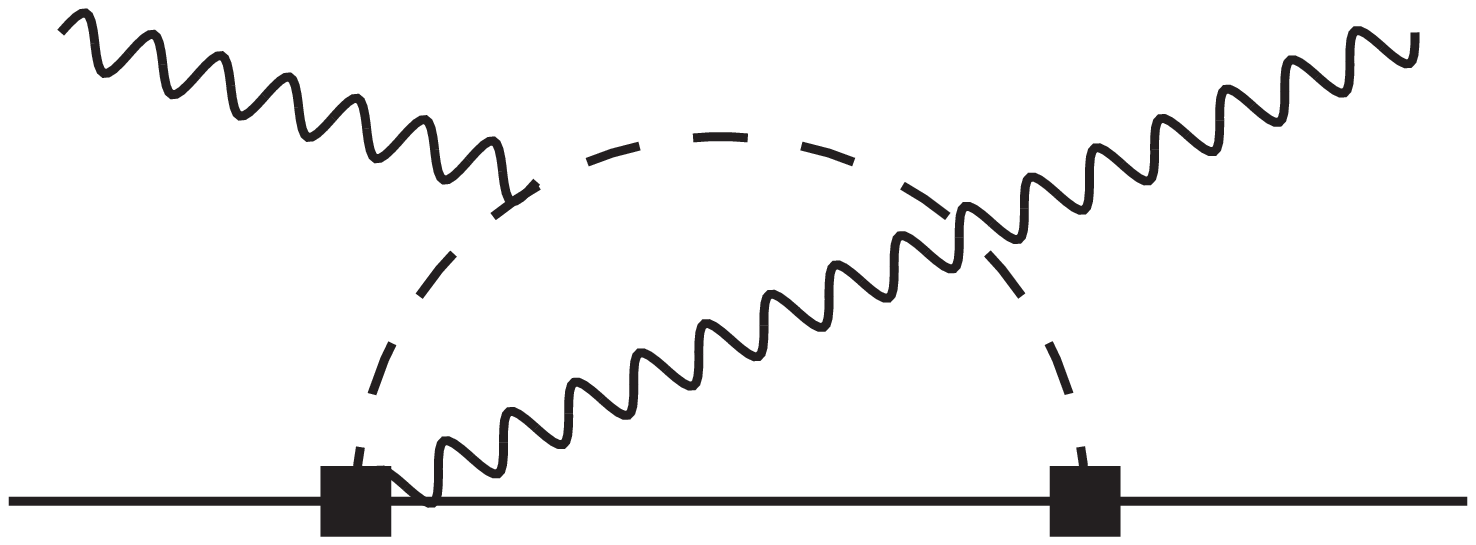} 
        & \includegraphics[width=0.3\columnwidth]{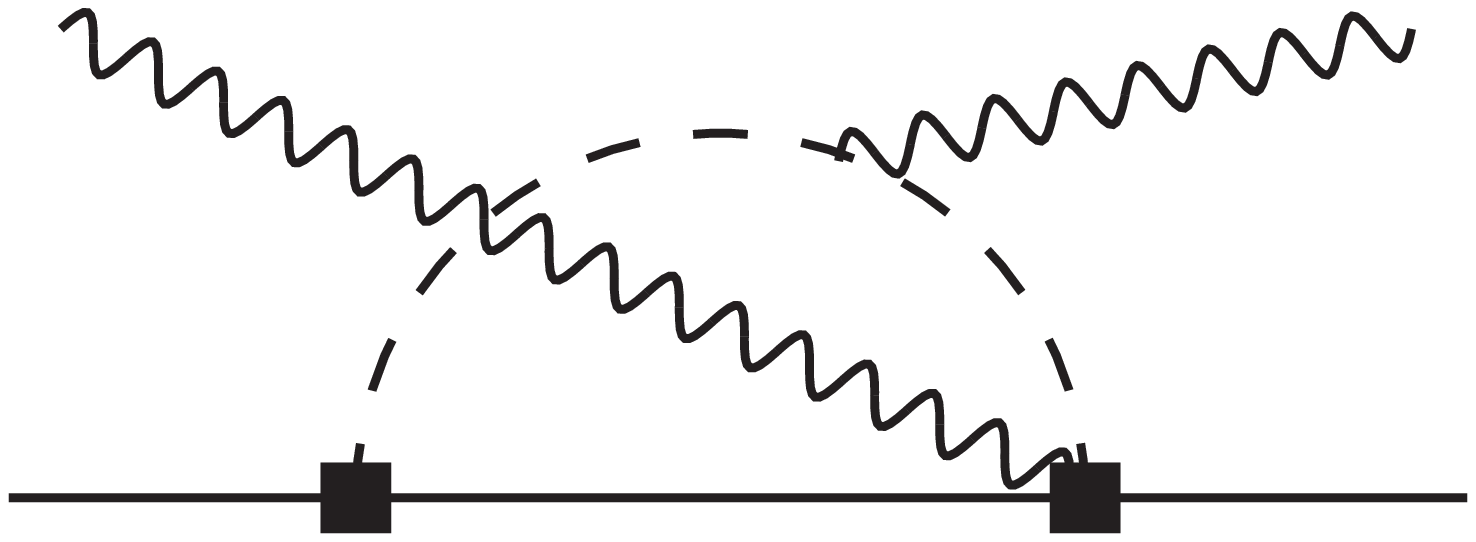} \\
        (d) & (e) &(f) \\ \\
        \includegraphics[width=0.3\columnwidth]{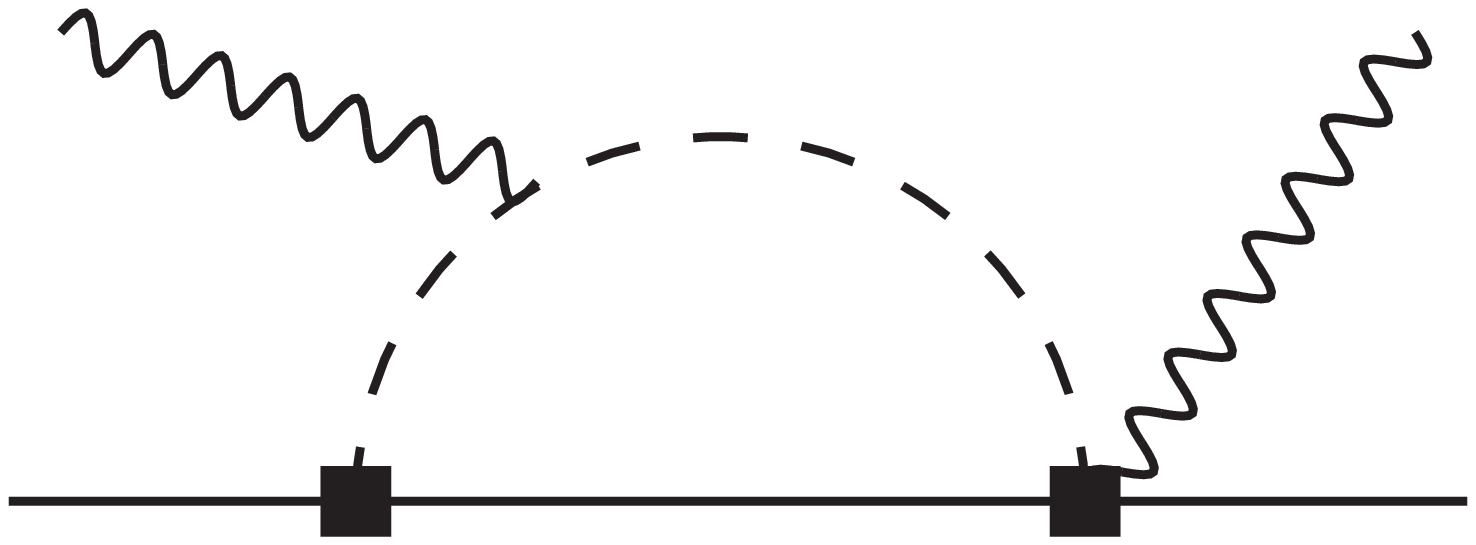} 
        & \includegraphics[width=0.3\columnwidth]{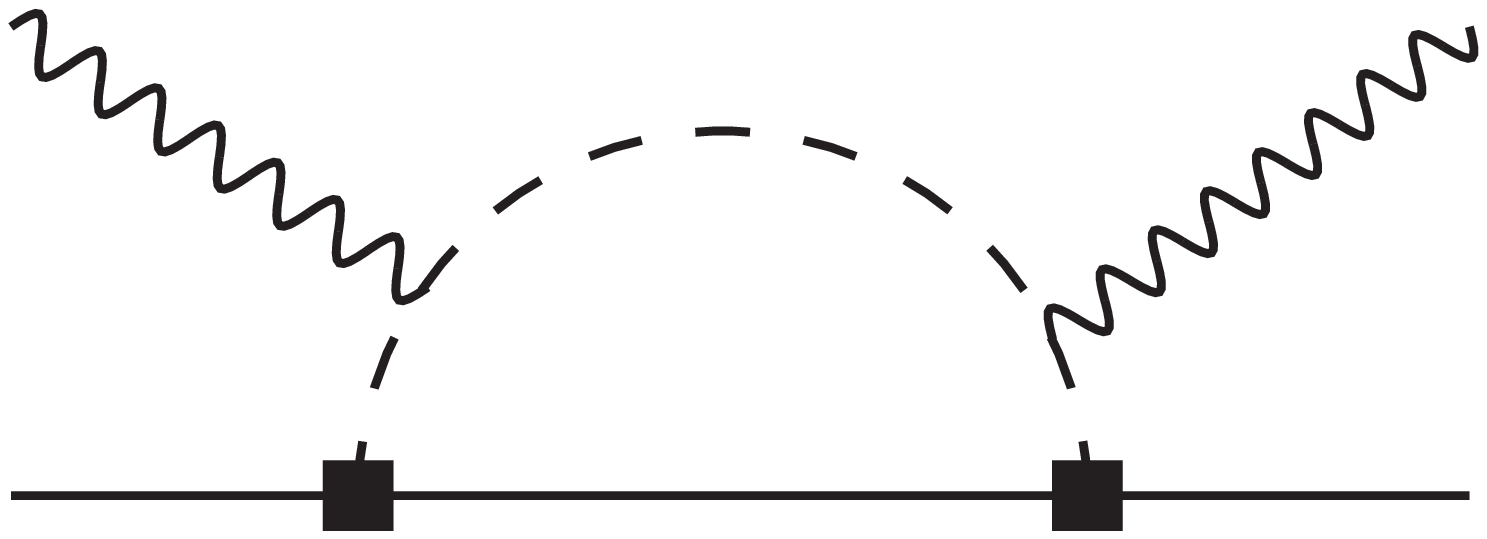} 
        &\includegraphics[width=0.3\columnwidth]{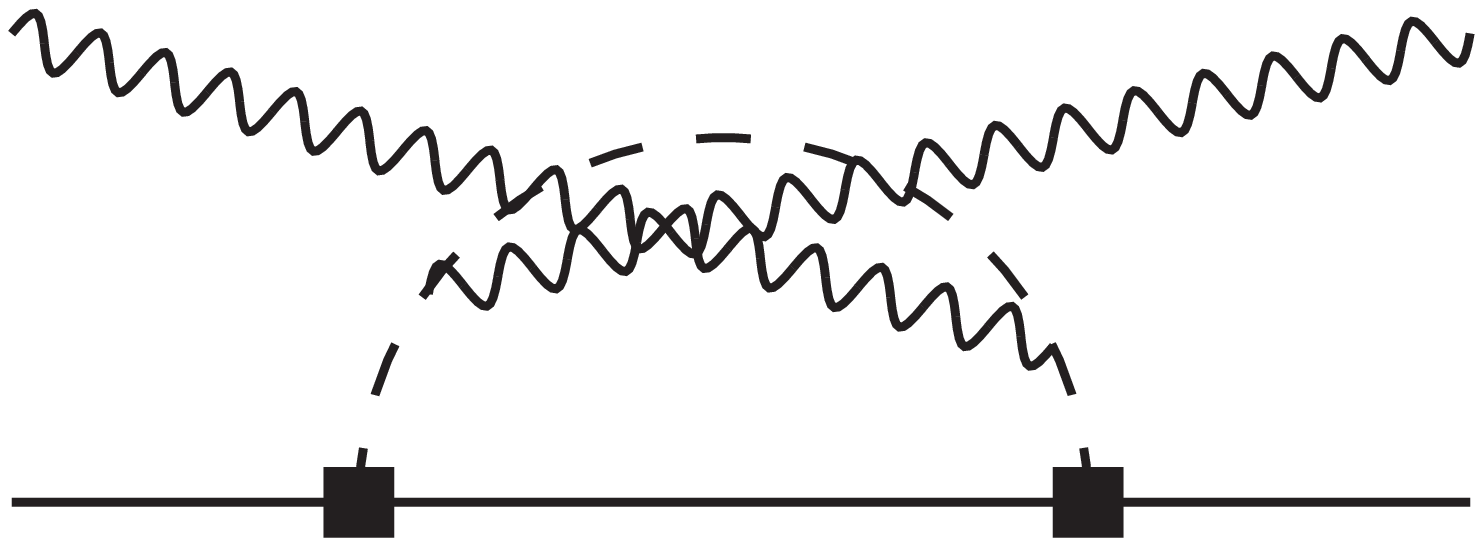} \\
        (g)& (h) & (i)
\end{tabular}
\caption{Diagrams contributing to nucleon polarisabilities at order
  $Q^3$. The solid and dashed lines denote nucleons, and pions
  respectively. Additionally, our results include a similar set of
  diagrams in which the internal {\bf 70}-plet propagator is replaced
  by a {\bf 44}-plet resonance.}
\label{fig:Q3}
\end{figure}

Using the Lagrangian of the preceding section, we can calculate the
amplitudes defined in Eq.~(\ref{eq:Ti}) for Compton scattering from a
nucleon (extensions to full octet and decuplet of baryons are
straight-forward although the convergence of \hbxpt\ with three-flavours
is not clear).  We work with a power counting such that
\begin{equation}
        Q \sim e \sim \frac{|\vec{p}|}{\L_\c} 
                \sim \frac{m_\pi}{\Lambda_\chi} 
                \sim \frac{\w}{\L_\c}\,
\end{equation}
(it is also convenient to count $\Delta/\Lambda_\c$ as the same as $Q$
as it is numerically similar at the masses relevant for current
lattice calculations).\footnote{Loop and pole~\cite{Butler:1992pn}
  contributions with {\bf 44}-plet intermediate states must be
  included since $\Delta$ is a small-scale. 
  Any $\Delta$ dependent terms analytic in $m_\pi$ arising from the
  loop diagrams, and additional operators proportional to powers of
  $\Delta/\L_\c$ can be resummed into the appropriate LECs of $\Delta$
  independent operators (the LECs then depend on
  $\Delta$)~\cite{Tiburzi:2004kd,Tiburzi:2005na}.  Keeping these
  contributions explicit is redundant as $\Delta$ can not be varied in
  a controlled manner.}  Below, we will also restrict ourselves to the
low frequency limit $\w \ll m_\pi$ in order to extract the
polarisabilities from the Compton scattering amplitudes defined in
Eqs.~(\ref{eq:Ti}) and (\ref{eq:amplitudes}).  For larger energies,
the concept of polarisabilities breaks down and the target essentially
becomes a dispersive medium. Working to order $Q^3$ in the chiral
expansion, Compton scattering requires the calculation of the diagrams
shown in Figs.~\ref{fig:anom}, \ref{fig:Born} and \ref{fig:Q3} (and a
corresponding set involving internal {\bf 44}-plet baryons). By
definition, tree level contributions from nucleon pole diagrams do not
contribute to the polarisabilities; their contribution to the
amplitudes are given explicitly in Eq.~(\ref{eq:amplitudes}).  For
each polarisability $X=\alpha,\,\beta,\,\gamma_{1\mbox{--}4}$, it is
convenient to separate the different contributions as
\begin{eqnarray}
  \label{eq:5}
  X= X^{\rm anomaly} + X^\Delta +X^{\rm loop} \,,
\end{eqnarray}
corresponding to the contributions from Figs \ref{fig:anom},
\ref{fig:Born} and \ref{fig:Q3}, respectively. We discuss these
contributions in the following subsections. At order $Q^3$, all
contributions are expressible in terms of a small set of LECs that
contribute in many other processes and are thus reasonably well
determined (at least in the \xpt\ case).  The total ${\cal O}(Q^3)$
loop contribution is finite, but loop-contributions at higher orders
are divergent; as discussed in the preceding section, the
counter-terms specific to Compton scattering that absorb these
divergences and the associated scale dependence enter at ${\cal
  O}(Q^4)$ for the electric and magnetic polarisabilities and ${\cal
  O}(Q^5)$ for the spin polarisabilities.

\subsection{Volume independent contributions to polarisabilities}
\label{sec:VolumeIndependent}

%
%
%
%
%
%
%
%
%
\subsubsection{Anomalous contribution to $\g N\rightarrow \g N$:
  $\pi^0 \rightarrow \g\g$}\label{sec:Anomaly} 

The anomalous decay of flavour neutral mesons to two photons
\cite{Bell:1969ts,Adler:1969gk} has important consequences in Compton
scattering in non-forward directions. These contributions arise from
the meson pole diagram shown in Fig.~\ref{fig:anom}.  Anomalous decays
are well understood in \CPT, entering through the Wess-Zumino-Witten
(WZW) Lagrangian \cite{Wess:1971yu,Witten:1983tw}.  However, these
effects have not been investigated in the quenched and partially
quenched theories and some interesting subtleties arise.

In SU(2) \CPT, the one-pion, two-photon piece of the WZW Lagrangian,
is given by
\begin{eqnarray}
        {\cal L}_{\pi^0 \g\g} &=& - \frac{3 e^2}{16 \pi^2 f} 
                \tr \left[\phi \hat{\cal Q}^2\right]
                \epsilon^{\mu\nu\rho\sigma}F_{\mu\nu}F_{\rho\sigma}, 
\label{eq:LWZW}
\end{eqnarray}
where
\begin{equation}
        \phi = \begin{pmatrix}
                        \frac{\pi^0}{\sqrt{2}} & \pi^+ \\
                        \pi^- & -\frac{\pi^0}{\sqrt{2}}
                \end{pmatrix}\quad, \quad
        \hat {\cal Q} = \begin{pmatrix}
                        \frac{2}{3} & 0 \\
                        0 & -\frac{1}{3}
                \end{pmatrix}.
\end{equation}
This Lagrangian is completely determined as its coefficient can be
fixed by directly matching to the perturbative QCD calculation of the
relevant triangle diagram (the one loop calculation is exact
\cite{Adler:1969er}, in accordance with Witten's geometric
quantisation condition \cite{Witten:1983tw}).  At higher orders,
additional anomalous operators appear \cite{Bijnens:2001bb} but they
do not contribute to Compton scattering until ${\cal O}(Q^5)$.

It is well known that quenched and partially-quenched chiral
perturbation theories generally have more complicated operator
structure than in the case of QCD ({\it e.g.}, one can not use
Cayley-Hamilton identities \cite{Sharpe:2003vy}). Thus, in order to
generalise Eq.~(\ref{eq:LWZW}) to the partially-quenched cases, we
might imagine the extended $\pi^0\to\g\g$ Lagrangian to be of the form
\begin{equation}
\label{eq:PQWZWwrong}
        {\cal L}_{\pi^0 \g\g}^{PQ} \propto 
                \epsilon^{\mu\nu\rho\sigma}F_{\mu\nu}F_{\rho\sigma}
                \Big[ 
                        a_1\, \str \left[ \Phi {\cal Q}^2 \right]  
                        +a_2\, \str \left[ \Phi {\cal Q} \right] \str
                        \left[ {\cal Q} \right]  
                        +a_3\, \str \left[ \Phi \right] \str
                          \left[ {\cal Q} \right] ^2  
                        +a_4\, \str \left[ \Phi \right] \str \left[
                          {\cal Q}^2 \right] \Big], 
\end{equation}
(in the quenched case only the first operator is non-vanishing, but a
similar discussion applies).  With the condition that in the QCD limit
where the sea-quark and ghost-quark masses and charges are set equal
to those of the valence-quarks, matrix elements of
Eq.~\eqref{eq:PQWZWwrong} reproduce the matrix elements of
Eq.~\eqref{eq:LWZW}. As discussed in Section~\ref{sec:hbxpt}, the
singlet field, $\Phi_0$ acquires a large mass from the strong U(1)$_A$
anomaly \cite{Sharpe:2000bn} and can be integrated out of the
partially-quenched theory; consequently, the operators proportional to
$a_3$ and $a_4$ can be ignored.  Additionally, from the multiple
super-trace structure, one can deduce that the operators
\begin{equation*}
        \str \left[ \Phi {\cal Q} \right] \str \left[ {\cal Q}
        \right], 
\quad
\str \left[ \Phi \right] \str \left[ {\cal Q} \right]^2, 
\quad 
\str \left[ \Phi \right] \str \left[
                          {\cal Q}^2 \right]\,,
\end{equation*}
have at least two closed loops at the quark level.  Following the
arguments in Refs.~\cite{Bell:1969ts,Adler:1969gk,Adler:1969er}, one
can show that these operators do not correspond to anomalous quark
level processes.  Moreover, the leading dependence of the underlying
quark-level diagrams is proportional to the quark mass, and thus the
coefficient of these operators must scale as, $a_{2,3,4} \sim
m_q/\Lambda_\chi^2$.  Although these operators contribute to $\eta_a
\to \g\g$, they are not anomalous, and only contribute at higher
orders in the chiral expansion.  We can thus conclude that the only
operator in the Lagrangian, Eq.~\eqref{eq:PQWZWwrong}, which
contributes to the anomalous decay of the neutral mesons at leading
order is $\str\left[\Phi{\cal Q}^2\right]$. The coefficient is easily
determined by matching to either perturbative partially-quenched QCD
or to the \CPT\ expression in the QCD limit.\footnote{We can thereby
  bypass the need to extend Witten's global quantisation condition
  \cite{Witten:1983tw} to non-compact graded Lie groups.}  The
appropriate Lagrangian is therefore
\begin{eqnarray}
        {\cal L}_{\pi^0 \g\g}^{PQ} &=& - \frac{3 e^2}{16 \pi^2 f} 
                \str \left[\Phi {\cal Q}^2\right]
                \epsilon^{\mu\nu\rho\sigma}F_{\mu\nu}F_{\rho\sigma}\,. 
\label{eq:LWZWPQ}
\end{eqnarray}

From the above Lagrangian, it is apparent that all of the flavour
diagonal fields in Eq.~(\ref{eq:Phi_def}), have anomalous couplings to
two photons. Calculating the diagrams in Fig~\ref{fig:anom} leads to
the following anomalous contribution to Compton scattering on the
proton\footnote{The anomalous contribution to neutron-Compton
  scattering is given by Eq.~\eqref{eq:TAnom} with the interchange of
  $u \leftrightarrow d$.} in partially-quenched \CPT
\begin{multline}\label{eq:TAnom}
        T^{PQ,{\rm anomaly}}_{\mu\nu} = -i\ \epsilon_{\mu\nu\alpha\beta}
       \, k^{\prime\alpha}\,k^\beta\, r \cdot S\, \frac{24 e^2}{(4\pi f)^2}
                  \Bigg\{
                        2 g_A \bigg[
                        \left(q_u^2-\frac{1}{2}q_j^2-\frac{1}{2}q_l^2
                        \right) \frac{1}{r^2-m_{uu}^2}
                        +\frac{\left( q_j^2-q_l^2 \right)
                        }{4}\frac{\D_{lj}^2}{(r^2-m_{uu}^2)(r^2-m_X^2)}\bigg] \\ 
                 +g_1 \bigg[ \left( q_u^2 -\frac{1}{2}q_j^2
                   -\frac{1}{2}q_l^2 \right) \frac{1}{r^2-m_{uu}^2} 
                        -\left( \frac{1}{2}q_j^2
                          +\frac{1}{2}q_l^2-q_d^2 \right)
                        \frac{1}{r^2-m_{dd}^2} \\ 
                        -\frac{ \left( q_j^2-q_l^2 \right)}{4} 
                                \frac{\D_{lj}^2}{(r^2-m_X^2)} \left(
                                  \frac{1}{(r^2-m_{uu}^2)} 
                                                +\frac{1}{(r^2-m_{dd}^2)} \right) 
                \bigg] \Bigg\}.
\end{multline}
In the above expression, $r=q^\prime-q$, is the momentum transfer to
the nucleon and $\D_{lj}^2 = m_{ll}^2-m_{jj}^2$ is a measure of the
isospin breaking in the sea sector.  In the sea isospin limit ($m_l
\to m_j$), the double pole structure of the amplitude vanishes, and in
the QCD limit all dependence on $g_1$ vanishes.

Expanding Eq.~(\ref{eq:TAnom}) in frequency and comparing with
Eq.~(\ref{eq:Ti}) leads to the following anomalous contributions to
the polarisabilities:
\begin{eqnarray}
        \alpha^{\rm anomaly}&=& 0\,,\\
        \beta^{\rm anomaly}&=& 0\,,\\
        \gamma_1^{\rm anomaly}&=& -\frac{3e^2G_{\rm anom}}{8\pi^3
                f^2m_\pi^2}\,, \\
        \gamma_2^{\rm anomaly}&=& 0\,,\\
        \gamma_3^{\rm anomaly}&=& \frac{3e^2G_{\rm anom}}{16\pi^3
                f^2m_\pi^2} \,,\\
        \gamma_4^{\rm anomaly}&=& -\frac{3e^2G_{\rm anom}}{16\pi^3
                f^2m_\pi^2} \,,
\end{eqnarray}
where the coefficients, $G_{\rm anom}$, are given in
Table~\ref{tab:couplings} for the different theories under
consideration.  These contributions vanish in the iso-scalar
combination of proton and neutron targets in the QCD limit.

\subsubsection{$\Delta$ resonance contributions}
\label{sec:Deltares}
The contributions to the amplitude from the Born-terms involving the
{\bf 44}-plet resonance (which contains the $\Delta$-isobar),
Fig.~\ref{fig:Born}, are identical in \CPT, \PQCPT\ and \QCPT\ as they
are purely valence quark processes. They are given by
\begin{eqnarray}
\alpha^\Delta &=& 0\,, \\
\beta^\Delta &=&   \mu_T^2\frac{e^2(q_u-q_d)^2}{36 \pi ( 2 M_N)^2 \Delta} \,,\\
\gamma_1^\Delta &=& 0 \,,\\
\gamma_2^\Delta &=&  -  \mu_T^2\frac{e^2(q_u-q_d)^2}{72 \pi ( 2 M_N)^2 \Delta^2} \,,\\
\gamma_3^\Delta &=& 0 \,,\\
\gamma_4^\Delta &=&  \mu_T^2\frac{e^2(q_u-q_d)^2}{72 \pi ( 2 M_N)^2 \Delta^2}\,,
\end{eqnarray}
where $\mu_T$ is the magnetic dipole transition coupling of
Eq.~(\ref{L1}).

%
%
%
%
%
%
%
%

\subsection{Infinite volume}
\label{sec:inf_vol}

The loop contributions to infinite volume chiral expansion of the
polarisabilities in \xpt\ are well known at order $Q^3$
\cite{Bernard:1991rq,Bernard:1991ru,Butler:1992ci,Hemmert:1996rw,Hemmert:1997tj,Pascalutsa:2002pi,Pascalutsa:2003zk}
and at ${\cal O}(Q^4)$
\cite{Bernard:1993ry,Bernard:1993bg,Ji:1999sv,VijayaKumar:2000pv,Gellas:2000mx,McGovern:2001dd,Beane:2002wn,Beane:2004ra}
(at this order, the $\Delta$-resonances have not been included as
dynamical degrees of freedom, restricting the range of applicability
to $m_\pi\ll\Delta$). Since the photon only couples to charged mesons,
the results in the quenched and partially-quenched theories are
similar to those in \xpt.  In particular, no quenched or
partially-quenched sicknesses (double pole contributions from neutral
meson propagators) enter expression for the loop diagrams.  In
general, the quenched power counting presents differences for
electromagnetic observables
\cite{Savage:2001dy,Arndt:2003vd,Arndt:2003ww,Arndt:2003we}, however
no new contributions appear at the order we work.

Using the effective couplings $G_B$, $G_B^\prime$, $G_T$ and
$G_T^\prime$ given in Table~\ref{tab:couplings}, we find that the loop
contributions to the polarisabilities are
\begin{table}[!t]
  \centering
  \begin{ruledtabular}
    \begin{tabular}{cccc}
            & QCD & QQCD & PQQCD \\ \hline
        $G_{\rm anom}$ 
        & $g_A\, (2Z-1)(q_u^2-q_d^2)$ & $2 g_A \left( Z q_u^2+ (1-Z) q_d^2 \right)$
        & $g_A \Big[ 2\Big( Z q_u^2+ (1-Z) q_d^2 \Big) -q_j^2-q_l^2\Big]$ 
\\
        & & $+g_1 \left(q_d^2+q_u^2\right)$ 
        & $+g_1\left(q_d^2-q_j^2-q_l^2+q_u^2\right)$    
\\\\ 
      $G_B$ & $4g_A^2\left(q_d-q_u\right)^2$ 
      & $\frac{1}{3}\left(4g_A^2-4g_A g_1-5g_1^2\right)\left(q_d-q_u\right)^2$ 
      & $-\frac{1}{3} \left(5 g_1^2+4 g_A g_1-4
        g_A^2\right) \left(q_d-q_u\right)^2$
\\ \\
      $G_B^\prime$ & 0 & 0 
      & $\begin{array}{l}
        \frac{1}{3} \Big[\left(6 q_d^2-6
          \left(q_j+q_l\right) q_d+5 q_j^2+5 q_l^2+4 q_u^2-4 q_j q_u-4
          q_l q_u\right) g_1^2 \\
        \hspace*{5mm}+4 g_A \left(q_j^2-2 q_u 
          q_j+q_l^2+2 q_u^2-2 q_l q_u\right) g_1 \\
        \hspace*{5mm}+8 g_A^2 \left(q_j^2-2 q_u
          q_j+q_l^2+2 q_u^2-2 q_l q_u\right)\Big]\end{array}$ 
\\ \\
      $G_T$ 
      & $\frac{4}{3}g_{\Delta N}^2\left(q_d-q_u\right)^2$
      & $\frac{5}{6}g_{\text{$\Delta $N}}^2\left(q_d-q_u\right)^2$
      & $\frac{5}{6}
      g_{\text{$\Delta $N}}^2 \left(q_d-q_u\right)^2$ 
\\ \\
      $G_T^\prime$ &  0 & 0 & $\frac{1}{6} g_{\text{$\Delta $N}}^2
      \left(4 q_d^2-4 \left(q_j+q_l\right) q_d+3 q_j^2+3 q_l^2+2
        q_u^2-2 q_j q_u-2 q_l q_u\right)$
%
   \end{tabular}
    \end{ruledtabular}
    \caption{Effective couplings for the various contributions to the polarisabilities.}
  \label{tab:couplings}
\end{table}
\begin{eqnarray}
  \label{eq:alpha_inf}
  \alpha^{\rm loop}&=&
\frac{e^2}{4\pi f^2}\left[\frac{5 G_B}{192 \pi}\frac{1}{m_\pi} 
+\frac{5 G_B^\prime}{192 \pi}\frac{1}{m_{uj}} 
+\frac{G_T}{72 \pi^2} F_\alpha(m_\pi,\D)
+\frac{G_T^\prime}{72 \pi^2}F_\alpha(m_{uj},\D)
\right]\,,
\end{eqnarray}
\begin{eqnarray}
  \label{eq:beta_inf}
  \beta^{\rm loop}&=&
\frac{e^2}{4\pi f^2}\left[\frac{ G_B}{384 \pi}\frac{1}{m_\pi} 
+\frac{G_B^\prime}{384 \pi}\frac{1}{m_{uj}} 
+\frac{G_T}{72 \pi^2}F_\beta(m_\pi,\D)
+\frac{G_T^\prime}{72 \pi^2}F_\beta(m_{uj},\D)
\right]\,,
\end{eqnarray}
\begin{eqnarray}
  \label{eq:gamma1_inf}
  \gamma_1^{\rm loop}&=&
\frac{e^2}{4\pi f^2}\left[\frac{ G_B}{48 \pi^2}\frac{1}{m_\pi^2} 
+\frac{G_B^\prime}{48 \pi^2}\frac{1}{m_{uj}^2} 
+\frac{G_T}{72 \pi^2} F_{\gamma_1}(m_\pi,\D)
+\frac{G_T^\prime}{72 \pi^2}F_{\gamma_1}(m_{uj},\D)
\right]\,,
\end{eqnarray}
\begin{eqnarray}
  \label{eq:gamma2_inf}
  \gamma_2^{\rm loop}&=&
\frac{e^2}{4\pi f^2}\left[\frac{ G_B}{96 \pi^2}\frac{1}{m_\pi^2} 
+\frac{G_B^\prime}{96 \pi^2}\frac{1}{m_{uj}^2} 
+\frac{G_T}{72 \pi^2} F_{\gamma_2}(m_\pi,\D)
+\frac{G_T^\prime}{72 \pi^2}F_{\gamma_2}(m_{uj},\D)
\right]\,,
\end{eqnarray}
\begin{eqnarray}
  \label{eq:gamma3_inf}
  \gamma_3^{\rm loop}&=&
\frac{e^2}{4\pi f^2}\left[\frac{ G_B}{192 \pi^2}\frac{1}{m_\pi^2} 
+\frac{G_B^\prime}{192 \pi^2}\frac{1}{m_{uj}^2} 
+\frac{G_T}{144 \pi^2} F_{\gamma_3}(m_\pi,\D)
+\frac{G_T^\prime}{144 \pi^2}F_{\gamma_3}(m_{uj},\D)
\right]\,,
\end{eqnarray}
\begin{eqnarray}
  \label{eq:gamma4_inf}
  \gamma_4^{\rm loop}&=&
-\frac{e^2}{4\pi f^2}\left[\frac{ G_B}{192 \pi^2}\frac{1}{m_\pi^2} 
+\frac{G_B^\prime}{192 \pi^2}\frac{1}{m_{uj}^2} 
+\frac{G_T}{144 \pi^2} F_{\gamma_4}(m_\pi,\D)
+\frac{G_T^\prime}{144 \pi^2}F_{\gamma_4}(m_{uj},\D)
\right]\,,
\end{eqnarray}
where
\begin{eqnarray}
  \label{eq:8}
  F_\alpha(m,\D) &=& \frac{9\Delta}{\Delta^2-m^2}-\frac{\Delta^2-10
    m^2}{2 (\Delta^2-m^2)^{3/2}}\ln \left[\frac{\Delta -
      \sqrt{\Delta^2 - m^2 + i \epsilon}}{\Delta + 
  \sqrt{\Delta^2 - m^2 + i \epsilon}}\right] \,, \\
  F_\beta(m,\D) &=& - \frac{1}{2(\Delta^2-m^2)^{1/2}}\ln \left[\frac{\Delta -
      \sqrt{\Delta^2 - m^2 + i \epsilon}}{\Delta + 
  \sqrt{\Delta^2 - m^2 + i \epsilon}}\right]\,, \\
F_{\gamma_1}(m,\D) &=& - \frac{\Delta^2 + 2 m^2}{(\Delta^2 - m^2)^2} -
\frac{3 \Delta m^2}{2 (\Delta^2 - m^2)^{5/2}} \ln \left[\frac{\Delta -
      \sqrt{\Delta^2 - m^2 + i \epsilon}}{\Delta + 
  \sqrt{\Delta^2 - m^2 + i \epsilon}}\right] \,,\\
F_{\gamma_2}(m,\D) &=& F_{\gamma_3}(m,\D) = F_{\gamma_4}(m,\D) =
\frac{1}{\Delta^2 - m^2} + \frac{\Delta}{2 (\Delta^2 - m^2)^{3/2}} \ln
\left[\frac{\Delta -
      \sqrt{\Delta^2 - m^2 + i \epsilon}}{\Delta + 
  \sqrt{\Delta^2 - m^2 + i \epsilon}}\right] \,.
\end{eqnarray}
Here we have used dimensional regularisation, however the results are
finite and hence independent of the regulator without the addition of
counterterms. These loop contributions vanish at zero photon
frequency, preserving the Thompson limit. They are identical for both
proton and neutron targets, though isospin breaking effects from
loops enter at ${\cal O}(Q^4)$ in the expansion.  In the \xpt\ case,
these results reproduce those of
Refs.~\cite{Hemmert:1996rw,Hemmert:1997tj}.

%
%
%
%
%
%
%
%
%
%
%
%
\subsection{Finite volume}
\label{sec:p_regime}

In momentum space, the finite volume of a lattice simulation restricts
the available momentum modes and consequently the results differ from
their infinite volume values. These long-distance effects can be
accounted for in the low-energy effective theory. Here we shall
consider a hyper-cubic box of dimensions $L^3\times T$ with $T\gg L$.
Imposing periodic boundary conditions on mesonic fields leads to
quantised momenta $k=(k_0,{\vec k})$, ${\vec k}=\frac{2\pi}{L} {\vec
  j}=\frac{2\pi}{L} (j_1,j_2,j_3)$ with $j_i\in \mathbb{Z}$, but $k_0$
treated as continuous.  On such a finite volume, spatial momentum
integrals are replaced by sums over the available momentum modes. This
leads to modifications of the infinite volume results presented in the
previous section; the various functions arising from loop integrals
are replaced by their finite volume (FV) counterparts. In a system
where $m_\pi L\gg 1$, the power counting of the infinite volume
low-energy effective theory remains valid and finite volume effects
are predominantly from Goldstone mesons propagating to large distances
where they are sensitive to boundary conditions and can even ``wrap
around the world''. Smaller volumes in which $m_\pi L\sim1$ are
discussed in Appendix~\ref{epp_regime}. Since the lowest momentum mode
of the Goldstone propagator is $\sim \exp(-m_\pi L)$ in position
space, finite volume effects will behave as a polynomial in $1/L$
times this exponential if no multi-particle thresholds are reached in
the loop (as is the case in these calculations provided the photon
energy is small enough, $\omega\lesssim m_\pi$).

Repeating the calculation of the loop diagrams using finite volume
sums rather than integrals leads to the following expressions for the
loop contributions to the polarisabilities:
\begin{eqnarray}
  \label{eq:alpha_fv}
  \alpha^{\rm loop}(L) &=& \frac{e^2}{1152 \pi f^2} \int_0^\infty
d\lambda\Big[
3G_B {\cal F}_\alpha({\cal M}_{uu})
+ 3G_B^\prime {\cal F}_\alpha({\cal M}_{uj})
+ 8G_T {\cal F}_\alpha({\cal M}_{uu}^\Delta)
+ 8G_T^\prime {\cal F}_\alpha({\cal M}_{uj}^\Delta)
\Big]\,,
\end{eqnarray}
\begin{eqnarray}
  \label{eq:beta_fv}
  \beta^{\rm loop}(L) &=& \frac{e^2}{1152 \pi f^2 }\int_0^\infty
d\lambda\Big[
3G_B {\cal F}_\beta({\cal M}_{uu})
+ 3G_B^\prime {\cal F}_\beta({\cal M}_{uj})
+ 8G_T {\cal F}_\beta({\cal M}_{uu}^\Delta)
+ 8G_T^\prime {\cal F}_\beta({\cal M}_{uj}^\Delta)
\Big]\,,
\end{eqnarray}
\begin{eqnarray}
  \label{eq:gamma1_fv}
  \gamma_1^{\rm loop}(L) &=&  \frac{7e^2}{576 \pi  f^2}\int_0^\infty
d\lambda\Big[
3G_B {\cal F}_{\gamma_1}({\cal M}_{uu})
+ 3G_B^\prime {\cal F}_{\gamma_1}({\cal M}_{uj})
-4G_T {\cal F}_{\gamma_1}({\cal M}_{uu}^\Delta)
-4 G_T^\prime {\cal F}_{\gamma_1}({\cal M}_{uj}^\Delta)
\Big]\,,
\end{eqnarray}
\begin{eqnarray}
  \label{eq:gamma2_fv}
  \gamma_2^{\rm loop}(L) &=&  \frac{7^2}{64 \pi f^2}\int_0^\infty
d\lambda\Big[
3G_B {\cal F}_{\gamma_2}({\cal M}_{uu})
+ 3G_B^\prime {\cal F}_{\gamma_2}({\cal M}_{uj})
-4 G_T {\cal F}_{\gamma_2}({\cal M}_{uu}^\Delta)
-4 G_T^\prime {\cal F}_{\gamma_2}({\cal M}_{uj}^\Delta)
\Big]\,,
\end{eqnarray}
\begin{eqnarray}
  \label{eq:gamma3_fv}
  \gamma_3^{\rm loop}(L) &=& \frac{7e^2}{1152 \pi f^2}\int_0^\infty
d\lambda\Big[
3G_B {\cal F}_{\gamma_3}({\cal M}_{uu})
+3 G_B^\prime {\cal F}_{\gamma_3}({\cal M}_{uj})
-4 G_T {\cal F}_{\gamma_3}({\cal M}_{uu}^\Delta)
-4 G_T^\prime {\cal F}_{\gamma_3}({\cal M}_{uj}^\Delta)
\Big]\,,
\end{eqnarray}
\begin{eqnarray}
  \label{eq:gamma4_fv}
  \gamma_4^{\rm loop}(L) &=& -\gamma_3^{\rm loop}(L)\,,
\end{eqnarray}
where ${\cal M}_{ab}=\sqrt{m_{ab}^2+\lambda^2}$ and ${\cal
  M}_{ab}^\Delta=\sqrt{m_{ab}^2+2\lambda \Delta +\lambda^2}$ and
\begin{eqnarray}
  \label{eq:F_alpha}
  {\cal F}_\alpha(m) &=& 
180 \lambda ^2 {\cal I}_{\frac{7}{2}}(m)+190
{\cal J}_{\frac{7}{2}}(m)-280 \lambda ^2 {\cal J}_{\frac{9}{2}}(m)-455
{\cal K}_{\frac{9}{2}}(m)+315 
   \lambda ^2 {\cal K}_{\frac{11}{2}}(m)+252 {\cal
L}_{\frac{11}{2}}(m) \,,
\\
  \label{eq:F_beta}
  {\cal F}_\beta(m) &=& 60 {\cal J}_{\frac{7}{2}}(m)-224
  {\cal K}_{\frac{9}{2}}(m)+189 {\cal L}_{\frac{11}{2}}(m) \,,
\\
  \label{eq:F_gamma1}
  {\cal F}_{\gamma_1}(m) &=& 30 \lambda ^3
  {\cal I}_{\frac{9}{2}}(m)+10 \lambda {\cal J}_{\frac{9}{2}}(m)-45
\lambda ^3
  {\cal J}_{\frac{11}{2}}(m)-18 \lambda{\cal K}_{\frac{11}{2}}(m)  \,,
\\
  \label{eq:F_gamma2}
  {\cal F}_{\gamma_2}(m) &=&\lambda{\cal K}_{\frac{11}{2}}(m) \,,
\\
  \label{eq:F_gamma34}
  {\cal F}_{\gamma_3}(m) &=& 10\lambda
  {\cal J}_{\frac{9}{2}}(m)-9 \lambda{\cal K}_{\frac{11}{2}}(m)  \,,
\end{eqnarray}
and the finite volume sums ${\cal I}_\beta(m)$, \ldots, ${\cal
  L}_\beta(m)$ are defined in Appendix \ref{FV_app}. These 
expressions reduce to the results of
Eqs.~(\ref{eq:alpha_inf})--(\ref{eq:gamma4_inf}) above in the infinite
volume limit. 

\begin{figure}[!th]
\centering
        \includegraphics[width=\columnwidth]{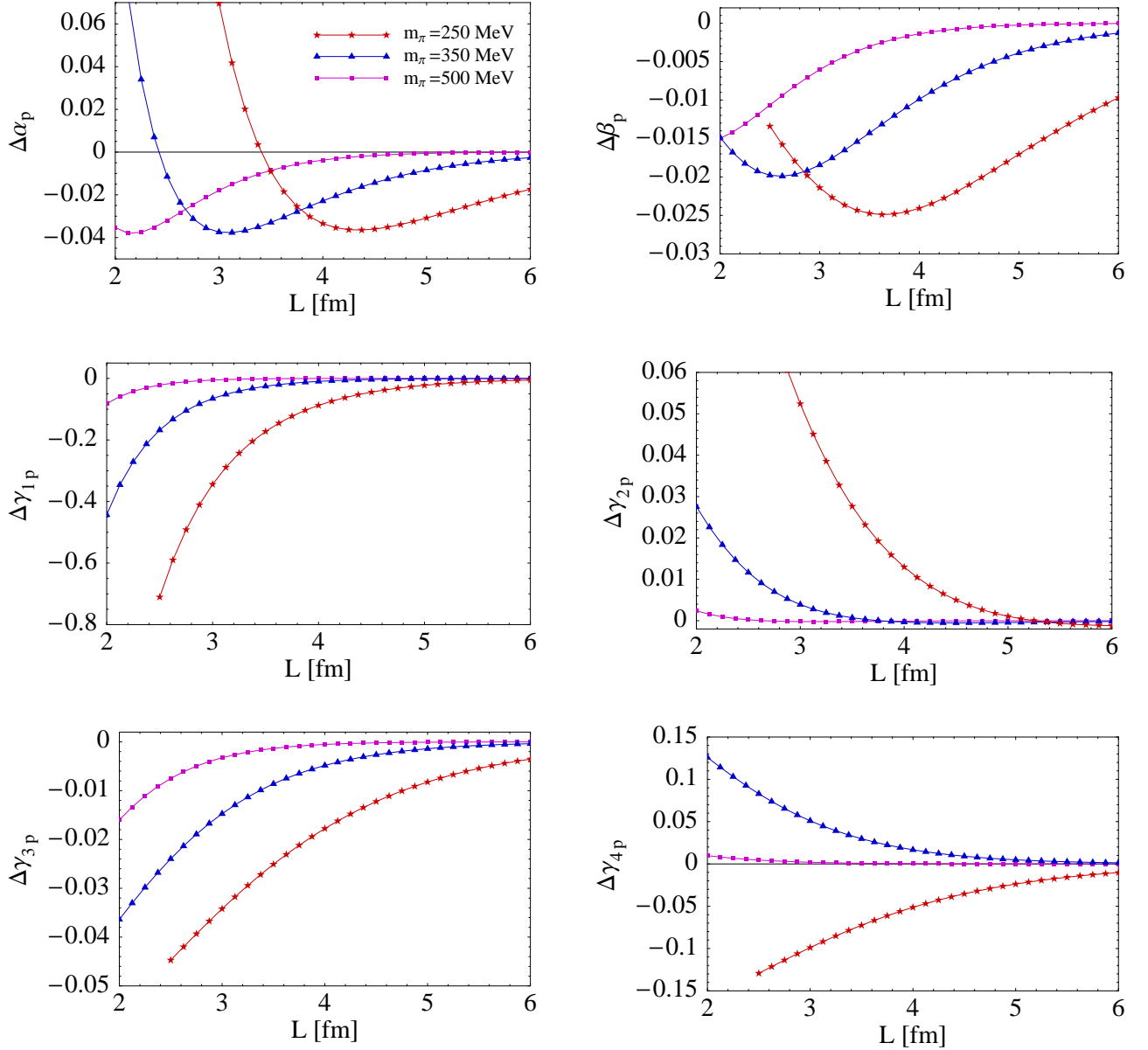}
\caption{Volume dependence of the proton polarisabilities. Here we
  show the ratio of the difference of the finite and infinite volume
  results to the infinite volume results for three values of the pion
  mass using the parameters described in the text. The curves
  terminate at $m_\pi\ L=3$.}
\label{fig:FV1}
\end{figure}

\begin{figure}[!th]
\centering
        \includegraphics[width=\columnwidth]{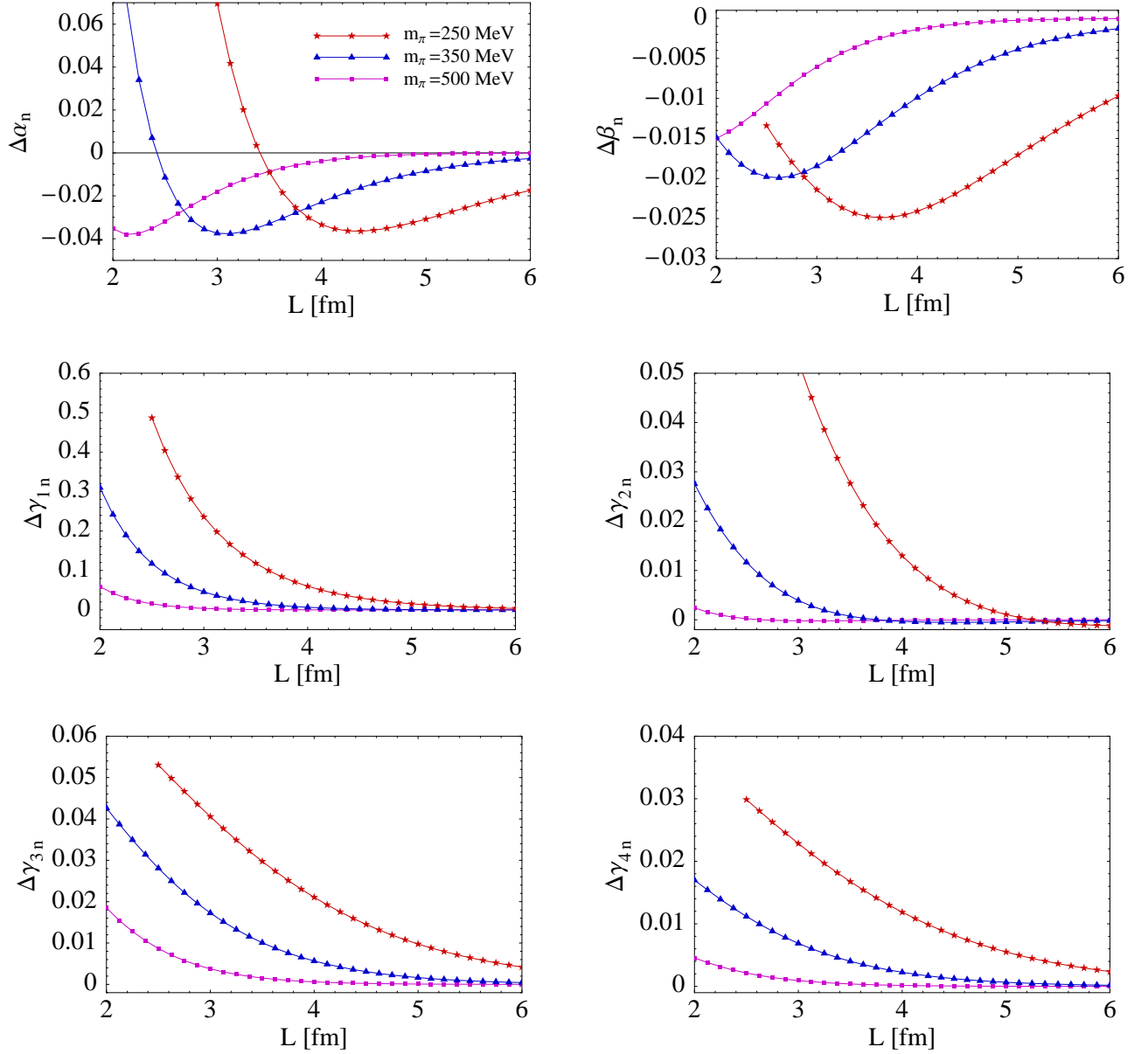}
\caption{Volume dependence of the neutron polarisabilities. The
  various curves are as in Fig.~\protect\ref{fig:FV1}. }
\label{fig:FV2}
\end{figure}

To illustrate these effects, Figs. \ref{fig:FV1} and \ref{fig:FV2}
show the volume dependence of the various polarisabilities in the
proton and the neutron, respectively. Here we have specialised to QCD,
setting $q_u=2/3$, $q_d=-1/3$, $g_A=1.25$, $|g_{N\Delta}|=1.5$,
$\mu_T=10.9$, $f=0.132$~GeV, $M_N=0.938$~GeV and
$\Delta=0.3$~GeV.%
\footnote{The value of $\mu_T$ is chosen to
  correspond to that found in analysis of Ref.~\cite{Hemmert:1997tj}
  ($\mu_T=2\sqrt{2}b_1$ of that reference).  In principle this LEC can
  be determined from an analysis of lattice polarisabilities or
  $N$--$\Delta$ transition matrix elements.}
In each plot we show results for the ratio
\begin{equation}
  \Delta X(L)=\frac{X(L)-X(\infty)}{X(\infty)}\,,
\end{equation}
for the six polarisabilities at three different pion masses,
$m_\pi=0.25,\,0.35,\,0.50$~GeV. The overall magnitude of these shifts
varies considerably; generally volume effects are at the level of
5--10\% for $m_\pi=0.25$~GeV and smaller for larger masses. Larger
effects are seen in a number of the spin polarisabilities but there
are as yet no lattice calculations of these quantities. The magnetic
polarisability has a particularly small volume dependence which can be
understood from the large decuplet resonance contribution that is
independent of the volume.

\begin{figure}[!t]
\centering
        \includegraphics[width=\columnwidth]{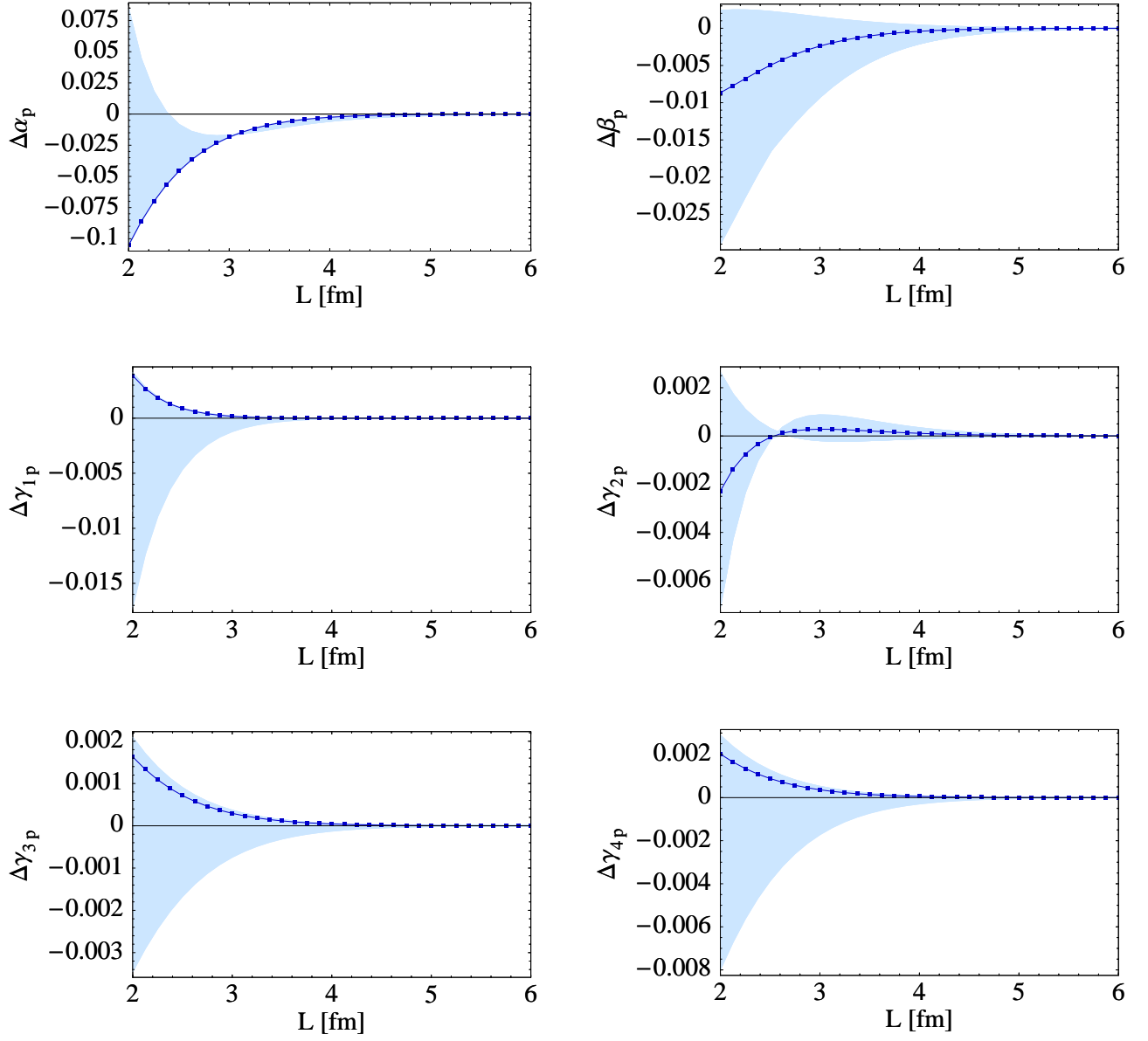}
\caption{Volume dependence of the proton polarisabilities in quenched
  QCD at the lightest quark mass used in the lattice calculations of
  Refs.~\protect{\cite{Christensen:2004ca,Lee:2005dq}}. The central
  curves and shaded region correspond to the parameters quoted in the
  text.}
\label{fig:FVQ_p}
\end{figure}

\begin{figure}[!t]
\centering
        \includegraphics[width=\columnwidth]{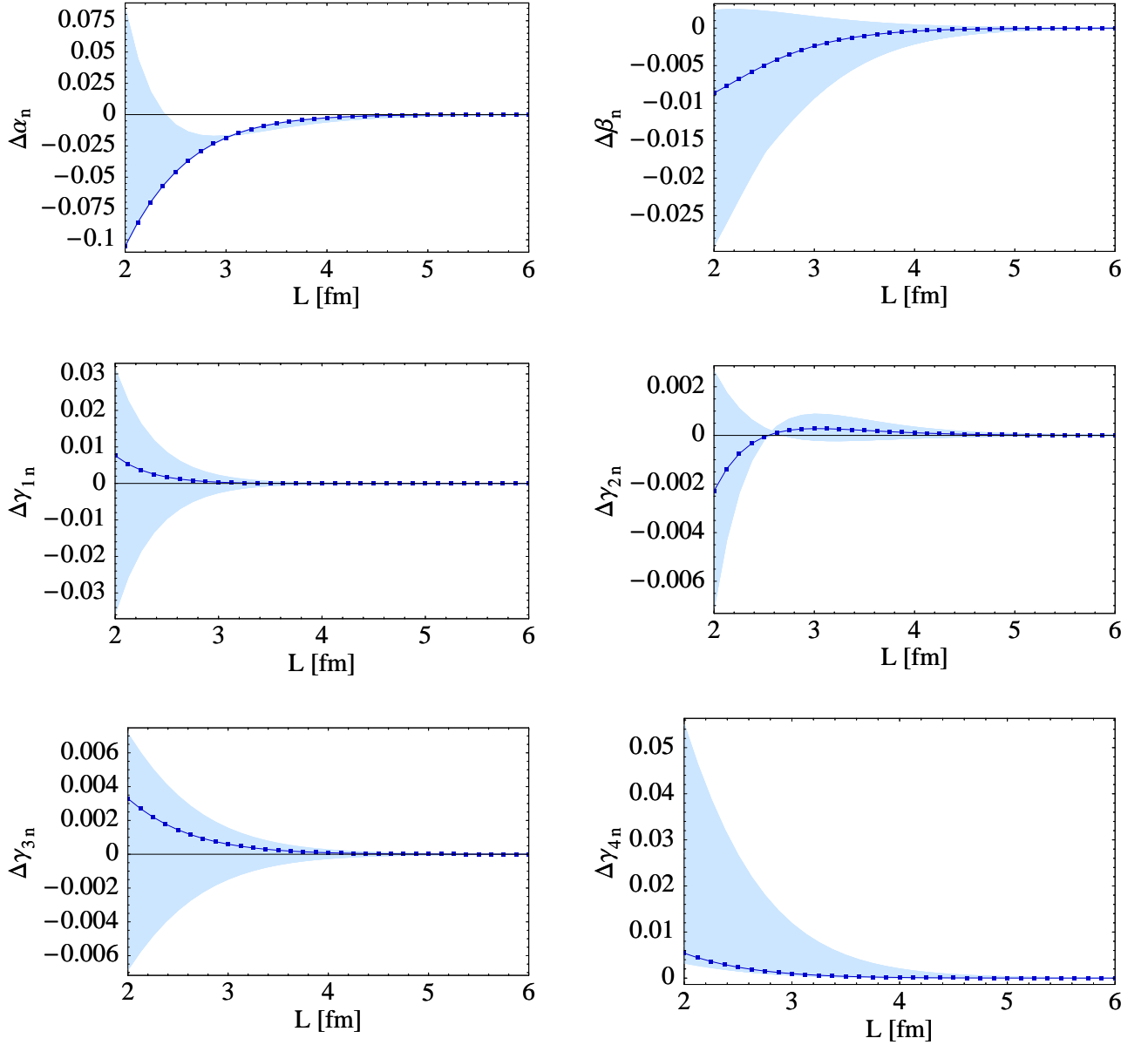}
\caption{As in Fig.~\protect\ref{fig:FVQ_p}, but for the neutron.}
\label{fig:FVQ_n}
\end{figure}

The above expressions also allow us to calculate the finite volume
effects in the quenched data on the various polarisabilities
calculated in Refs.~\cite{Christensen:2004ca,Lee:2005dq}. The quenched
expressions involve a number of undetermined LECs (quenched $g_A$,
$g_1$, $g_{N\Delta}$ and $\mu_T$ are unrelated to their \pqxpt/\xpt\ 
values), so we can only estimate the volume effects. To do so, we
choose $q_u=2/3$, $q_d=-1/3$, $g_A=1.25$, $g_1=1$,
$|g_{N\Delta}|=1.5$, $\mu_T=5.85$, $f=0.132$~GeV, $M_N=0.938$~GeV and
$\Delta=0.3$~GeV and take the pion masses corresponding to the
lightest used in these lattice calculations, $m_\pi\sim 0.5$~GeV (we
ignore the issue of the convergence of \xpt\ at such masses). The
results for the volume dependence of the various polarisabilities of
the proton and neutron are shown in Figs.~\ref{fig:FVQ_p} and
\ref{fig:FVQ_n}. In each plot, the shaded region corresponds to
reasonable variation of the unknown couplings, $-1<g_1<1$,
$0.8<|g_{N\Delta}|<2$ and $2.8<|\mu_T|<8.5$.  From these figures, we
see that the calculations on a (2.4 fm)$^3$ lattice with $m_\pi=$0.5~GeV
may differ from their infinite volume values by 5--10\% in the case of
the electric polarisability and a few percent for the magnetic and
spin polarisabilities.

\begin{figure}[!th]
\centering
        \includegraphics[width=0.75\columnwidth]{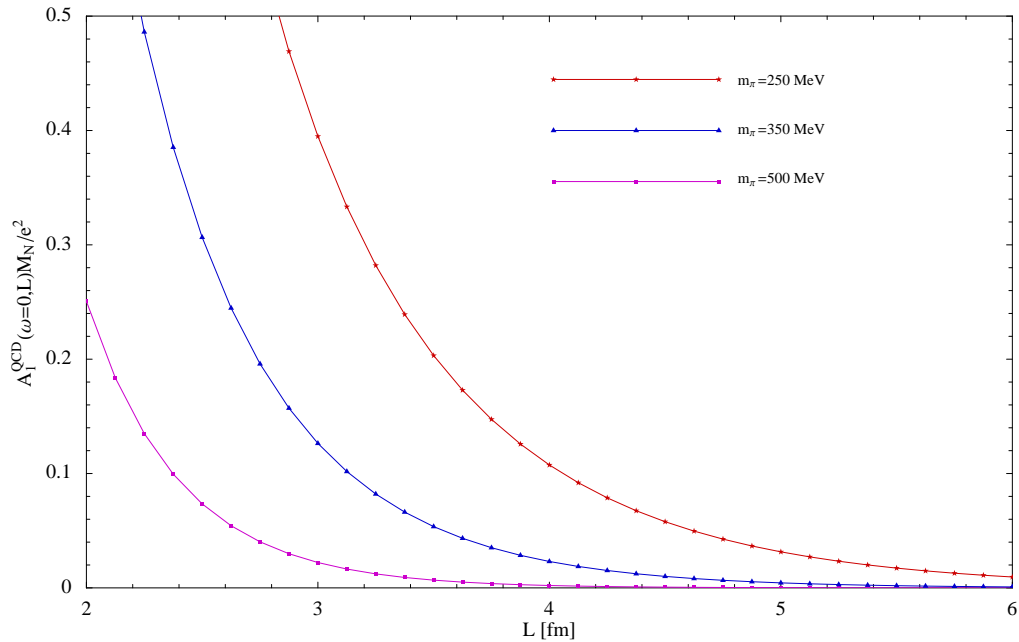}
\caption{Volume dependence of the Thompson limit of photon neutron
scattering. Notice the infinite volume limit is zero. }
\label{fig:FVThomson}
\end{figure}

An interesting effect that arises at finite volume is that the
Thompson-limit and other Born terms in the frequency expansion of the
scattering amplitude (terms in Eq.~(\ref{eq:amplitudes}) that are not
polarisabilities) receive finite volume contributions from the loop
diagrams in Fig.~\ref{fig:Q3} that vanish exponentially as the volume
is increased. As an example, the amplitude for Thomson-limit (zero
frequency) scattering on the neutron (which is identically zero at
infinite volume) is shown in Fig~\ref{fig:FVThomson}. This result is
somewhat counter-intuitive, but arises from the effects of the
periodic boundary conditions on the long range charge distribution of
the hadron. It does not imply the non-conservation of charge.

The results presented here all assume that the higher order terms in
the $Q$ expansion provide small contributions to the volume dependence
of the polarisabilities. This may or may not be the case as diagrams
that are formally of higher order in the infinite volume \xpt\ 
power-counting can have volume effects that are enhanced over those at
lower infinite volume order (see Ref.~\cite{Detmold:2005pt} for a
detailed discussion). Such issues may be particularly relevant for the
polarisabilities where the convergence of the chiral expansion is
tenuous. In this regard, studying the FV behaviour of the lattice
results may in fact be a useful diagnostic tool with which to
determine if or why the convergence is poor.

%
%
%
%
%
%
%
%
%
%
%
%
\section{Conclusion}
\label{sec:discussion}

We have investigated Compton scattering from spin-half targets from
the point of view of lattice QCD. We first discussed how external
field methods can be used to probe all six polarisabilities of real
Compton scattering for both charged and uncharged targets. Such
calculations will tell us a lot about the low energy QCD structure of
hadrons and will be of much use in phenomenological studies requiring
the full set of polarisabilities as only certain linear combinations
are available from current
experiments~\cite{Schumacher:2005an,Hyde-Wright:2004gh}. The
techniques discussed here also allow us to extract other electric
properties of charged particles using external fields including the
electric dipole moment of the proton and the quadrupole moment of the
deuteron.

Our second major focus was on the effects of the finite volume used in
lattice calculations on the polarisabilities. Since polarisabilities
are infrared-sensitive observables (they scale as inverse powers of
the pion mass near the chiral limit), the are expected to have strong
volume dependence. This is indeed borne out in the explicit
calculations presented here. In QCD, we generically find that the
polarisabilities experience volume shifts of 5--10\% from the infinite
volume values for lattice volumes $\sim$(2.4~fm)$^3$ and pions of mass
0.25~GeV. The electric and first spin polarisabilities are
particularly sensitive. In the case of quenched QCD (relevant to the
only existing lattice data), we find significant effects even at pion
masses $\sim0.5$~GeV. Future lattice studies of the polarisabilities
should take these effects into account in order to present physically
relevant results.

As extensions of this work, one can also consider the generalised
polarisabilities, higher-order polarisabilities and parity violating
polarisabilities (see Ref.~\cite{Bedaque:1999dh}) all of which can be
extracted from appropriate lattice calculations similar to those
detailed in Section~\ref{sec:Compton}. Such information would lead to
a further-improved understanding of the low-energy structure of the
hadrons and prove very useful in directing the next generation of
precision Compton scattering experiments. The lattice provides a novel
opportunity to study the neutron polarisabilities directly instead of
from nuclear targets and extending the lattice methods of Section
\ref{sec:Compton} to the deuteron (along similar lines to those
discussed in Ref.~\cite{Detmold:2004qn}) will also prove useful for
comparison to experiment.

\acknowledgments We express out gratitude to C.-J.~D.~Lin,
D.~O'Connell, D.~R.~Phillips, M.~J.~Savage, S.~R.~Sharpe, R. M. Woloshyn and
R.~D.~Young for many helpful discussions.  This work is supported by
the US Department of Energy under contracts DE-FG02-97ER41014 (WD and
AW-L) and DE-FG02-05ER41368-0 (BCT).

%
%
%
%
%
%
%
%
%
%
%
%
\appendix

%
%
%
%
%
%
%
%
%
%
%
%
\section{Quenched chiral Lagrangian}
\label{A4}

In this Appendix, we display the relevant pieces of the quenched
chiral Lagrangian in the meson and baryon sectors and note particular
pathologies of the quenched theory.  In a quenched two flavor theory,
we have valence ($u$, $d$) and ghost ($\tilde u,\,\tilde d$) quarks
with masses contained in the matrix
\begin{equation}
\label{eq:Mq_def2}
\overline{m}_Q = {\rm  diag}(m_u,m_d,m_{\tilde u},m_{\tilde d})\,,
\end{equation}
where $m_{\tilde u,\tilde d}=m_{u,d}$ to maintain the exact
cancellation from the path-integral determinants arising from the
valence and ghost quark sectors.  The corresponding low-energy meson
dynamics are described by the \qxpt\ Lagrangian. At leading order, the
form of this Lagrangian is the same as in Eq.~\eqref{eq:PGBlagrangian}
where the pseudo-Goldstone mesons are embedded non-linearly in
$\Sigma$ with the matrix $\Phi$ now given by
\begin{equation}
  \label{eq:Phi_def2}
\Phi = \begin{pmatrix} M & \chi^\dagger \cr \chi &\tilde{M} \end{pmatrix}
\,,
\end{equation}
where
\begin{eqnarray}
  \label{eq:Mchi_def2}
  M=\begin{pmatrix} \eta_u    & \pi^+        \\
                    \pi^-     & \eta_d      
                    \end{pmatrix}\,, 
& 
\hspace*{15mm}
\tilde{M}=\begin{pmatrix} \tilde\eta_u    & \tilde\pi^+        \\
                          \tilde\pi^-     & \tilde\eta_d       
                          \end{pmatrix}\,, 
 \hspace*{15mm}
&
\chi=\begin{pmatrix} \chi_{\eta_u} & \chi_{\pi^+}       \\
                     \chi_{\pi^-}  & \chi_{\eta_d} 
                     \end{pmatrix}\,.
\end{eqnarray}
The matrix $M$ contains the usual valence--valence mesons, while
mesons in $\tilde{M}$ are composed of ghost quarks and anti-quarks,
and finally those in $\chi$ of ghost--valence quark--anti-quark pairs.
Unlike the partially quenched theory, there is no strong U(1)$_A$
anomaly, and the flavor-singlet field, $\Phi_0=\str[\Phi]/\sqrt{2}$
(along with its couplings $m_0$ and $\alpha_\Phi$), must be retained
in the theory. For the electromagnetic and spin polarisabilities in
QQCD, no loop contributions from the singlet are needed to the order
we work as flavor-neutral mesons are not present in loop diagrams at
this order. Despite flavour neutral mesons being absent in loop
graphs, the anomalous tree-level term couples the quenched singlet to
the nucleon. Cancellations, however, lead to final results that are
independent of $m_0$ and $\alpha_\Phi$.

For the quenched electric charge matrix of the valence and ghost
quarks, we choose
\begin{eqnarray}
  \label{eq:88}
  \hat{\cal Q} &=& {\rm diag}(q_u,q_d,q_u,q_d)\,.
\end{eqnarray}
Notice the peculiarity that $\str \, \hat{\cal Q} = 0$ is unavoidable
in the quenched theory.  In the quenched theory, there are anomalous
decays of flavour neutral mesons into two photons. In terms of
SU(2$|$2) QQCD quark fields, contributions to the anomaly from the
valence and ghost sectors come weighted with squares of the quark
charges, and we are thus not restricted to only the flavor singlet
current (as is the case for the strong U(1)$_A$ anomaly).  The
relevant term of the anomalous quenched chiral Lagrangian is the same
as has been detailed above in Sec.~\ref{sec:Anomaly}.

In SU(2$|$2) \hbxpt, the nucleons (those composed of three valence
quarks) enter as part of a {\bf 20}-dimensional representation
described by a three index flavour-tensor, ${\cal B}$.  The quenched
$\Delta$-isobar is contained in the totally symmetric three index
flavour-tensor ${\cal T}^\mu$ transforming in the {\bf 12}-dimensional
representation of SU(2$|$2).  The leading-order Lagrangian describing
these baryons and their interactions with Goldstone mesons is
\begin{eqnarray}
\label{eq:free_lagrangian2}
{\cal L}_{B\;Q}^{(0)} & = & 
i\left(\overline{\cal B} v\cdot {\cal D} {\cal B}\right)
+ 2\alpha \left(\overline{\cal B} S^\mu {\cal B} {\cal A}_\mu\right)
+ 2\beta  \left(\overline{\cal B} S^\mu {\cal A}_\mu {\cal B} \right)
+ 2\gamma \left(\overline{\cal B} S^\mu {\cal B} \right) \, \str \cA_\mu
\\
&& \hspace*{-3mm}
- i \left(\overline{\cal T}^\mu v\cdot {\cal D} {\cal T}_\mu\right) 
+ \Delta\ \left(\overline{\cal T}^\mu {\cal T}_\mu\right)
+  2{\cal H} \left(\overline{\cal T}^\nu S^\mu {\cal A}_\mu {\cal T}_\nu \right)
+  2 \gamma' \left(\overline{\cal T}^\nu S^\mu {\cal T}_\nu \right) \, \str \cA_\mu
+ \sqrt{\frac{3}{2}}{\cal C} 
\left[
\left( \overline{\cal T}^\nu {\cal A}_\nu {\cal B}\right) + 
\left(\overline{\cal B} {\cal A}_\nu {\cal T}^\nu\right) \right].
\notag
\end{eqnarray}
In contrast to partially-quenched and unquenched chiral perturbation
theory, there are two additional axial couplings $\gamma$ and
$\gamma'$ due to the presence of the flavour-singlet field. One should
keep in mind that although we use the same notation for simplicity,
all of the coefficients in the quenched Lagrangian have distinct
numerical values from those of the partially-quenched Lagrangian. In
the large $N_c$ limit, the coefficients of the two theories are
related~\cite{Chen:2002mj}.

Again the photon is minimally coupled in the above Lagrangian with
fixed coefficients. At the next order in the expansion, the relevant
terms that appear are
\begin{eqnarray}
{\cal L}_{B\;Q}^{(1)}& = &
\frac{i\,e}{ 2 M_N} F_{\mu\nu}
\left[ 
\mu_\alpha \left( \overline{\cal B} \left[S^{\mu},S^{\nu}\right]  {\cal B}  
{\cal Q}_{\xi+} \right) +
\mu_\beta \left( \overline{\cal B} \left[S^{\mu},S^{\nu}\right]
{\cal Q}_{\xi+} 
{\cal B} \right)
 \right]
 \nonumber \\
&&
+ \sqrt{\frac{3}{2}} \mu_T \frac{i e}{ 2 M_N} F_{\mu\nu}
\left[
\left( \overline{\cal B} S^\mu {\cal Q}_{\xi+} {\cal T}^\nu  
 \right) +
\left( \overline{\cal T} {}^\mu S^\nu  {\cal Q}_{\xi+} {\cal B} \right) 
\right]
.\label{L12}
\end{eqnarray}
The \pqxpt\ term with coefficient $\mu_\gamma$ is absent in the
quenched theory.  This only affects the Born terms of the Compton
amplitude, which are essentially unknown because they depend on the
quenched magnetic moment.  Finally, the leading two-photon operators
that give completely local contributions to the Compton scattering
tensor appear in quenched chiral perturbation theory in essentially
the same form as \pqxpt.
However, there are fewer operators per spin structure compared to the
partially quenched case because of the super-tracelessness of the
electric charge matrix.  Our computation is unchanged since these
terms do not contribute at the order we work.

%
%
%
%
%
%
%
%
%
%
%
%

\section{Compton scattering in small volumes}
\label{epp_regime}

In this appendix, we discuss the volume dependence of the
polarisabilities on asymmetric lattices in which the spatial
dimensions are small but the temporal direction remains large:
$m_\pi\,L\ll1$ but $L^3\,T m_q\langle\bar{q}q\rangle\gg
1$.\footnote{The calculation of the volume dependence of the
  polarisabilities in the $\epsilon$-regime \cite{Gasser:1987ah}
  (where $m_\pi\,L\sim m_\pi T \sim1$ and $L^3\,T
  m_q\langle\bar{q}q\rangle\sim 1$ and zero-modes become
  non-perturbative) is beyond the scope of this work.} In this
$\epsilon^\prime$-regime \cite{Detmold:2004ap},
$m_\pi\sim\epsilon^{\prime 2}$ and $L\sim\epsilon^{\prime -1}$ (where
$\epp$ is the small expansion parameter) so Goldstone boson zero modes
(modes with $\vec{q}=0$) are enhanced, but remain perturbative. Thus,
the power counting of loop diagrams involving zero-modes is modified.
In Compton scattering at finite volume, the photon momenta provide
additional scales whose power-counting must be specified. The spatial
components of the incoming and outgoing photon momenta are quantised
as $\vec{q}^{(\prime)}=\frac{2\pi}{L}\vec{n}_{q^{(\prime)}}$ (where
$\vec{n}_{q^{(\prime)}}$ are integer 3-tuples) and consequently scale
as ${\cal O}(\epp)$. For real Compton scattering,
$q_0^{(\prime)2}=|\vec{q}^{(\prime)}|^2$, implying that the frequency
$\omega$ is either zero or also ${\cal O}(\epp)$, parametrically
larger than the pion mass, $\omega/m_\pi\gg1$. This introduces an
intrinsic difficulty in defining polarisabilities for realistic pion
masses in small volumes, as they arise from an expansion around the
zero frequency limit. Thus our discussion is restricted to the full
amplitudes $A_i(\omega,\theta)$ in Eq.~(\ref{eq:Ti}).  This is not an
issue in larger volumes (Sec.~\ref{sec:p_regime} above) as the
quantisation of momenta is fine-grained on the scale of $m_\pi$.  In
(doubly) virtual Compton scattering, the on-shell condition is relaxed
and one can again consider non-zero energies comparable to or smaller
than the pion mass.

In the $\epp$-regime, the loop diagrams in Fig.~\ref{fig:Q3}
generically contribute at order $\epsilon^{\prime 3}$ for non-zero
modes (we count $e\sim\epp$). However, diagrams 3(a) an 3(b) contain
non-derivative couplings and zero-modes provide a further enhanced
contribution, $\sim\epp$. Although the remaining diagrams in
Fig.~\protect{\ref{fig:Q3}} contain derivatively coupled pions, the
non-zero momentum insertions allow the energy integral to be performed
with a pole $k_0\sim m_\pi\sim\epsilon^{\prime 2}$, leading to a
putative enhancement.  However, transversality of the Compton
amplitude causes these enhanced contributions to vanish.  Thus
$A_1(\omega=0)$ receives contributions at ${\cal O}(\epp)$ and $A_3$
at ${\cal O}(\epsilon^{\prime 2})$ and these amplitudes will exhibit
enhanced volume dependence for lattice calculations in the
$\epp$-regime. This dependence is given by
\begin{eqnarray}
  \label{eq:eppres1}
  A_1(\omega,\theta) &=& \frac{e^2G_B}{4m_\pi^2L^3} + \frac{2e^2G_T}{3m_\pi
    L^3}\frac{1}{m_\pi+\Delta}    + {\cal O}(\epsilon^{\prime 3}) \,,
 \\ 
  \label{eq:eppres3}
  A_3(\omega,\theta) &=& \frac{e^2G_B}{4m_\pi
    L^3}\frac{\omega}{m_\pi^2-\omega^2} - \frac{e^2G_T}{3m_\pi
    L^3}\frac{\omega}{(m_\pi+\Delta)^2-\omega^2} 
  + {\cal O}(\epsilon^{\prime 3})\,.
\end{eqnarray}
Since $m_\pi$ is necessarily smaller than any non-zero value of
$\omega$ in the small volume regimes, pions can go on-shell in the
loop diagrams of Fig.~\ref{fig:Q3}. Consequently the power-law
dependence on volume seen in Eqs.~(\ref{eq:eppres1}) and
(\ref{eq:eppres3}) is not unexpected. The other amplitudes have
contributions from the diagrams in Fig.~\ref{fig:Q3} starting at
${\cal O}(\epsilon^{\prime 3})$. At this order additional diagrams
such as those arising from tadpole dressings of the vertices in
Fig.~\ref{fig:Q3}(a,b) also contribute and the full results are left
to future work.

%
%
%
%
%
%
%
%
%
%
%
%
\section{Finite volume functions}
\label{FV_app}

The sums required in the evaluation of the polarisabilities at finite
volume are ($\vec{k}=\frac{2\pi\vec{n}}{L}$, with $\vec{n}$ a triplet
of integers)
\begin{eqnarray}
  \label{eq:7a}
  {\cal I}_{\beta}(M)&=&\frac{1}{L^3}\sum
  _{\vec{k}}\frac{1}{\left[|\vec{k}|^2+M^2\right]^\beta}\,,
\\
  {\cal J}_\beta(M)&=&{\cal I}_{\beta-1}(M)-M^2 {\cal I}_\beta(M)\,,
  \\
  {\cal K}_\beta(M) &=& {\cal I}_{\beta-2}(M)
  - 2M^2 {\cal  I}_{\beta-1}(M)+M^4{\cal I}_\beta(M)\,,
  \\
  {\cal L}_\beta(M) &=&{\cal I}_{\beta-3}(M)
  -3M^2 {\cal I}_{\beta-2}(M)+3M^4{\cal I}_{\beta-1}(M)
  -M^6{\cal I}_\beta(M) \,.
\end{eqnarray}
At infinite volume these can be simplified using,
\begin{eqnarray}
  \label{eq:10}
  {\cal I}_\beta(M,L\to\infty) =
  \frac{1}{(4\pi)^{\frac{3}{2}}} 
  \frac{\Gamma(\beta -\frac{3}{2})}{\Gamma(\beta)}
  \frac{1}{(M^2)^{\beta-\frac{3}{2}}}  \,,
\end{eqnarray}
for $\beta>3/2$.

In numerically evaluating these sums, it is useful to note that
\begin{eqnarray}
\sum_{\vec{n}}\frac{1}{(|\vec{n}|^2+x^2)^\beta}
&=&
\sum_{\vec{n}}\frac{E_{1-\beta}(|\vec{n}|^2+x^2)}{\Gamma(\beta)}+\frac{\pi^{\frac{3}{2}}}{\Gamma(\beta)}
\int_0^1dt\, t^{\beta-5/2}e^{-t\,
  x^2}\left[\sum_{\vec{n}\ne0}e^{-\frac{\pi^2 |\vec{n}|^2}{t}}+1\right]
\end{eqnarray}
where $E_n(x)$ is the exponential integral function. This form is
valid for $\beta>\frac{3}{2}$, $x\in\mathbb{R}$ and the remaining sums
converge exponentially fast in $|\vec{n}|$.

\bibliography{MasterBib,EM_bib}

\end{document}